\documentstyle[aps]{revtex}

\begin{document}
\title{Massive Gauge Field Theory Without Higgs Mechanism\\
II. Proof of Renormalizability}
\author{Jun-Chen Su}
\address{Center for Theoretical Physics, College of Physics,\\
Jilin University, Changchun 130023,\\
People's Republic of China}
\date{}
\maketitle

\begin{abstract}
~~It is shown that the quantum massive non-Abelian field theory established
in the former papers is renormalizable. This conclusion is achieved with the
aid of the Ward-Takahashi identities satisfied by the generating functionals
which were derived in the preceding paper based on the BRST-symmetry of the
theory. By the use of the Ward-Takahashi identity, it is proved that the
divergences occurring in the perturbative calculations for the massive gauge
field theory can be eliminated by introducing a finite number of
counterterms in the effective action. As a result of the proof, it is found
that the renormalization constants for the massive gauge field theory comply
with the same Slavnov-Taylor identity as that for the massless gauge field
theory. The latter identity is re-derived from the Ward-Takahashi identities
satisfied by the gluon proper vertices and their renormalization. To
illustrate the renormalizability of the theory, the renormalization of the
QCD with massive gluons is concretely performed and the one-loop effective
coupling constant of the QCD is derived and given an exact expression in the
renormalization.

\end{abstract}

\section{Introduction}

As was mentioned in our previous paper$^{[1]}$ , in the original attempts of
setting up the massive non-Abelian gauge field theory without Higgs mechanism%
$^{[2-10]}$, the massive Yang-Mills Lagrangian itself was considered to form
a complete formulation of the massive gauge field dynamics and used to
establish the quantum theory. In the theory, there are two problems which
were announced to be difficult to solve: one is the gauge-non--invariance of
the mass term in the action, another is the nonrenormalizability of the
quantum theory. In Ref. [1], the first problem has readily been circumvented
from the viewpoint that the massive gauge field only exists in the physical
space spanned by the transverse part of the vector potential. In this space,
the action of the massive non-Abelian gauge fields whose masses are taken to
be the same is gauge-invariant with respect to the infinitesimal gauge
transformations which are only necessary to be taken into account in the
physical space . If we want to represent the massive gauge field dynamics in
the whole space of the vector potential, the massive gauge field must be
viewed as a constrained system. The Lorentz gauge condition, acting as a
constraint, must be introduced initially and imposed on the Lagrangian
expressed by the full vector potential. From this point of view, it has been
shown that the massive gauge field theory can well be established on the
basis of gauge invariance. In the preceding paper (referred to as paper I
later on), It was shown that the quantum massive gauge field theory
described in Ref.[1] has a BRST-symmetry, that is to say, the effective
action and the generating functional of Green functions are invariant with
respect to a kind of BRST-transformations. From the BRST-symmetry of the
theory, a series of Ward-Takahashi (W-T) identities$^{[10-12]}$ obeyed by
the generating functionals were derived and used to prove the unitarity of
the theory.

In this paper, we are devoted to proving the renormalizability of the
quantum massive gauge field theory. The renormalizability of such a theory
may be seen from the intuitive observation that the Feynman rules derived
from the effective action, except for the gluon and ghost particle
propagators, are the same as those given in the massless gauge field theory,
and the massive propagators have the same ultraviolet behavior as the
massless ones. In particular, the primitively divergent diagrams are
completely the same in the both theories. This fact suggests that the power
counting argument of demonstrating the renormalizability is useful not only
for the massless gauge field theory, but also for the massive gauge field
theory. Theoretically, to accomplish a rigorous proof of the
renormalizability of the massive non-Abelian gauge field theory, it is
necessary to implement a subtraction procedure to see whether the
divergences occurring in perturbative calculations of Green functions and
S-matrix elements can be removed by introducing a finite number of
counterterms in the action . This procedure, as one knows, amounts to the
well-known R-operation invented by Bogoliubov, Parasiuk, Hepp and Zimmermann$%
^{[13-15]}$. The principal idea of proving the renormalizability of the
theory under consideration is the usage of the W-T identities derived in
paper I. In the Landau gauge, as mentioned before, these identities formally
are identical to those for the massless gauge field theory. Therefore, the
proof of the renormalizability almost is the same as for the massless
theory. From the proof, it will be seen that the divergences occurring in
perturbative calculations can surely be eliminated by introducing a finite
number of counterterms. As a consequence of the proof. it will be found that
the Slavnov-Taylor identity$^{[16,17]}$ for the renormalization constants
which was derived in the massless gauge field theory also holds for the
massive gauge field theory. As will be shown, the Slavnov-Taylor identity
can also be derived from the W-T identities satisfied by the vertices and
their renormalization. To illustrate concretely the renormalizability of the
theory, the one-loop renormalization of the quantum chromodynamicm with
massive gluons (will be called massive QCD) will be performed and the
one-loop effective coupling constant for the QCD will be derived and given
an exact expression. From this expression, the previous result for the QCD
with massless gluons which was obtained by the minimal subtraction$%
^{^{[18-20]}}$ will be recovered in the large momentum limit.

The remainder of this paper is arranged as follows. In section 2, we
describe the general proof of the renormalizability of the massive
non-Abelian gauge field theory. In the proof, we start from the generating
functional of proper vertices and use the W-T identity for this generating
functional to show that the divergences occurring in the perturbative
expansion of the generating functional can completely be eliminated by
introducing a finite number of counterterms in the action. The counterterms
are explicitly given by solving the equations obeyed by the counterterms
which are derived from the W-T identities. In section 3, we analyze the role
played by the counterterms and prove the equivalence between the additional
renormalization and the multiplicative renormalization. as a result of the
proof, the Slavnov-Taylor identity satisfied by the renormalization
constants is naturally derived. Section 4 is used to derive the W-T identity
satisfied by the gluon three-line proper vertex, discuss its renormalization
and show how the identity for the renormalization constants can follow from
the above-mentioned W-T identity. In section 5, the same things will be done
for the gluon four-line proper vertex. The last section serves to make some
concluding remarks. Section 6 serves to derive the one-loop effective
coupling constant for the QCD with massive gluons. In Appendix, we quote the
proof which shows how the counterterms written in section 2 satisfy the W-T
identities.

\section{W-T identity and counterterms}

From this section, we start to prove the renormalizability of the massive
non-Abelian field theory in which all the fields have the same masses. The
theory concerned typifies the quantum chromodynamics (QCD) with massive
gluons$^{[10]}$, the SU(2)-symmetric hadrodynamics$^2$ or some others. In
the following, we will take the QCD with massive gluons as an example. Since
all of vertices and even Green functions in the theory may be derived from
the generating functional $\Gamma $ for proper vertices which was defined in
Eq. (3.11) in paper I, it is only necessary to deal with the renormalization
of such a generating functional and to see whether the divergences appearing
in the perturbative expansion of the functional $\Gamma $ can be removed by
introducing a finite number of counterterms .

Let us make use of the loop diagram expansion, which is a power series in
the Planck constant $\hbar $, for the proper vertex generating functional$%
^{[10,11]}$ 
\begin{equation}
\hat \Gamma =\sum_{n=0}^\infty \hat \Gamma _n  \eqnum{2.1}
\end{equation}
where $\hat \Gamma $ is, according to Eq. (3.21) in paper I, defined by
excluding the gauge-fixing term 
\begin{equation}
\hat \Gamma =\Gamma +\frac 1{2\alpha }\int d^4x(\partial ^\mu A_\mu ^a)^2. 
\eqnum{2.2}
\end{equation}
In the tree diagram approximation, as one knows, the proper vertex
generating functional $\Gamma _0$ just is the generalized action which is
defined by the ordinary effective action plus the BRST-external source terms
as was shown in Eq. (3.4) in paper I and, according to the definition given
in Eq. (2.2), it will be rewritten as 
\begin{equation}
\Gamma _0\equiv S_0=\hat \Gamma _0-\frac 1{2\alpha }\int d^4x(\partial ^\mu
A_\mu ^a)^2  \eqnum{2.3}
\end{equation}
where 
\begin{equation}
\begin{array}{c}
\hat \Gamma _0\equiv \hat S_0=\int d^4x\{\bar \psi [i\gamma ^\mu (\partial
_\mu -igA_\mu ^aT^a)]\psi -M\bar \psi \psi -\frac 14F^{a\mu \nu }F_{\mu \nu
}^a \\ 
+\frac 12m^2A^{a\mu }A_\mu ^a+\bar C^a\partial ^\mu ({\cal D}_\mu ^{ab}C^b)+%
\bar \zeta \Delta \psi +\Delta \bar \psi \zeta +u^{a\mu }\Delta A_\mu ^a \\ 
+v^a\Delta C^a\}.
\end{array}
\eqnum{2.4}
\end{equation}
In the above, $A_\mu ^a$ represent the gauge fields, $\psi $ and $\bar \psi $
denote the quark fields, $m$ and $M$ are the gluon and quark masses
respectively, 
\begin{equation}
{\cal D}_\mu ^{ab}(x)=\frac{\mu ^2}{\Box _x}\partial _\mu ^x+D_\mu ^{ab}(x) 
\eqnum{2.5}
\end{equation}
here $\mu ^2=\alpha m^2$ and 
\begin{equation}
D_\mu ^{ab}=\delta ^{ab}\partial _\mu -gf^{abc}A_\mu ^c  \eqnum{2.6}
\end{equation}
which is the ordinary covariant derivative, $\Delta \psi $, $\Delta \bar \psi
$, $\Delta A_\mu ^a$ and $\Delta C^a$ are the composite field functions
which are defined from the BRST-transformations by dropping out the
infinitesimal Grassmann number $\xi $ and $u^{a\mu }$, $v^a$, $\overline{%
\zeta }$ and $\zeta $ are the corresponding BRST-sources.

It should be emphasized that according to the additional renormalization
scheme which is adopted in the latter proof, all the field functions and
parameters in the effective action shown in Eqs. (2.3) - (2.6) are
renormalized ones and hence finite. In perturbative calculations, each loop
term in Eq. (2.1) is divergent and therefore has to be regularized by an
appropriate regularization scheme which must be chosen to preserve the
BRST-symmetry of the theory. Suppose the $n$-th term $\hat \Gamma _n$ in Eq.
(2.1) has been separated into a finite part $\hat \Gamma _n^f$ and a
divergent part $\hat \Gamma _n^d$ through the regularization procedure 
\begin{equation}
\hat \Gamma _n=\hat \Gamma _n^f+\hat \Gamma _n^d  \eqnum{2.7}
\end{equation}
In perturbative calculations, the divergences included in Eq. (2.1) may be
eliminated order by order through a recursive construction of counterterms
in the action. For instance, to eliminate the one-loop divergence $\hat 
\Gamma _1^d$ which is generated by using the action shown in Eq. (2.4) in
the following perturbative expansion 
\begin{equation}
\hat \Gamma [\hat S_0]=\hat \Gamma _0[\hat S_0]+\hat \Gamma _1[\hat S%
_0]+\cdot \cdot \cdot \cdot \cdot =\hat S_0+\hat \Gamma _1^d[\hat S_0]+\hat 
\Gamma _1^f[\hat S_0]+\cdot \cdot \cdot \cdot \cdot \cdot ,  \eqnum{2.8}
\end{equation}
we may choose a counterterm $\Delta S_0$ such that 
\begin{equation}
\Delta \hat S_0=-\hat \Gamma _1^d[\hat S_0]  \eqnum{2.9}
\end{equation}
whose concrete form will be given later. It is apparent that when we use the
action 
\begin{equation}
\hat S_1=\hat S_0+\Delta \hat S_0  \eqnum{2.10}
\end{equation}
to recalculate the functional $\hat \Gamma $ of order $\hbar $%
\begin{equation}
\begin{array}{c}
\hat \Gamma [\hat S_1]=\hat \Gamma [\hat S_0+\Delta \hat S_0]=\hat \Gamma _0[%
\hat S_0+\Delta \hat S_0]+\hat \Gamma _1[\hat S_0+\Delta \hat S_0]+\cdot
\cdot \cdot \cdot \cdot \\ 
=\hat S_0+\Delta \hat S_0+\hat \Gamma _1^d[\hat S_0]+\hat \Gamma _1^f[\hat S%
_0]+\cdot \cdot \cdot \cdot \cdot \\ 
=\hat S_0+\hat \Gamma _1^f[\hat S_0]+\cdot \cdot \cdot \cdot \cdot ,
\end{array}
\eqnum{2.11}
\end{equation}
the divergence $\hat \Gamma _1^d[\hat S_0]$ in it disappears. In general, to
remove the divergent part $\hat \Gamma _n^d$ in the $n$-loop term $\hat 
\Gamma _n$ of order $\hbar ^n$, we need to introduce a counterterm like this 
\begin{equation}
\Delta \hat S_{n-1}=-\hat \Gamma _n^d.  \eqnum{2.12}
\end{equation}
Adding it to the action, we have 
\begin{equation}
\hat S_n=\hat S_{n-1}+\Delta \hat S_{n-1}  \eqnum{2.13}
\end{equation}
where the $\hat S_{n-1}$ has included the counterterms up to the order $%
\hbar ^{n-1}$. The action $\hat S_n$ used to calculate the $\hat \Gamma _n$
will lead to a finite result in the $n$-th order perturbation.

The counterterms mentioned above may be determined with the help of the
following W-T identity satisfied by the generating functional of proper
vertices which was derived in paper I 
\begin{equation}
\hat \Gamma *\hat \Gamma +\omega =0  \eqnum{2.14}
\end{equation}
and the ghost equation 
\begin{equation}
\frac{\delta \hat \Gamma }{\delta \bar C^a(x)}-\partial _x^\mu \frac{\delta 
\hat \Gamma }{\delta u^{a\mu }(x)}-{\mu }^2C^a(x)=0  \eqnum{2.15}
\end{equation}
where 
\begin{equation}
\hat \Gamma *\hat \Gamma =\int d^4x\{\frac{\delta \hat \Gamma }{\delta A_\mu
^a}\frac{\delta \hat \Gamma }{\delta u^{a\mu }}+\frac{\delta \hat \Gamma }{%
\delta C^a}\frac{\delta \hat \Gamma }{\delta v^a}+\frac{\delta \hat \Gamma }{%
\delta \psi }\frac{\delta \hat \Gamma }{\delta \bar \zeta }+\frac{\delta 
\hat \Gamma }{\delta \bar \psi }\frac{\delta \hat \Gamma }{\delta \zeta }\} 
\eqnum{2.16}
\end{equation}
and 
\begin{equation}
\omega =m^2\int d^4x\partial ^\mu A_\mu ^a(x)C^a(x)  \eqnum{2.17}
\end{equation}
As mentioned in paper I, in the Landau gauge, since $\mu =0$ and $\omega =0$
due to $\partial ^\mu A_\mu ^a(x)=0$, the above W-T identity and ghost
equation are identical to those for the massless gauge theory. In the loop
expansion, when Eqs. (2.1) and (2.7) are substituted in Eq. (2.14), one may
obtain a series of identities satisfied by the finite and divergent parts of
the $\hat \Gamma _n$. Each of the identities contains terms which are of the
same order of divergence. The identities of order ${\hbar }^n$ are$%
^{[10-11]} $%
\begin{equation}
\begin{array}{c}
\sum_{p+q=n}\hat \Gamma _p^f*\hat \Gamma _q^f+\omega \delta _{n0}=0, \\ 
\hat \Gamma _{n-1}^f*\hat \Gamma _1^d+\hat \Gamma _1^d*\hat \Gamma
_{n-1}^f=0, \\ 
\cdot \cdot \cdot \cdot \cdot \cdot \cdot \cdot \cdot \cdot \cdot \cdot \\ 
\hat \Gamma _1^f*\hat \Gamma _{n-1}^d+\hat \Gamma _{n-1}^d*\hat \Gamma
_1^f=0, \\ 
\rho (\hat S_0)\hat \Gamma _n^d+\sum_{p+q=n}\hat \Gamma _p^d*\hat \Gamma
_q^d=0
\end{array}
\eqnum{2.18}
\end{equation}
where 
\begin{equation}
\rho (\hat S_0)=\frac{\delta \hat S_0}{\delta \varphi _i}\cdot \frac \delta {%
\delta u_i}+\frac{\delta \hat S_0}{\delta u_i}\cdot \frac \delta {\delta
\varphi _i}  \eqnum{2.19}
\end{equation}
here $\varphi _i$ and $u_i(i=A,\bar \psi ,\psi ,C)$ stand for the field
variables $A_\mu ^a,\bar \psi ,\psi ,C^a$ and source variables $u_\mu
^a,\zeta ,\bar \zeta ,v^a$, respectively, and the symbol ''.'' in each term
on the right hand side (RHS) of Eq. (2.19) is an abbreviation notation of
the integration as shown in Eq. (2.16). In the above, the fact that the $%
\omega $ is finite and of zeroth order of ${\hbar }$ has been noticed. This
fact is obvious because we start from the renormalized field functions and
parameters in the additional renormalization.

Furthermore, the action constructed in Eq. (2.13) is also required to
fulfill the W-T identity 
\begin{equation}
\hat S_n*\hat S_n+\omega =0.  \eqnum{2.20}
\end{equation}
On substituting Eq .(2.13) into Eq. (2.20) and noticing Eq. (2.12), one may
find three equations$^{[10-12]}$ 
\begin{equation}
\rho (\hat S_{n-1})\hat \Gamma _n^d=0,  \eqnum{2.21}
\end{equation}

\begin{equation}
\hat \Gamma _n^d*\hat \Gamma _n^d=0  \eqnum{2.22}
\end{equation}
and 
\begin{equation}
\hat S_{n-1}*\hat S_{n-1}+\omega =0  \eqnum{2.23}
\end{equation}
In Eq. (2.21), the operator $\rho (\hat S_{n-1})$ is defined as 
\begin{equation}
\rho (\hat S_{n-1})=\frac{\delta \hat S_{n-1}}{\delta \varphi _i}\cdot \frac 
\delta {\delta u_i}+\frac{\delta \hat S_{n-1}}{\delta u_i}\cdot \frac \delta
{\delta \varphi _i}.  \eqnum{2.24}
\end{equation}
When we set the subscript $n=2,3,\cdot \cdot \cdot $ and repeatedly apply
Eqs. (2.12) and (2.13), it is easy to find that the second term on the left
hand side (LHS) of the last equation in Eq. (2.18) equals to zero. Thus, the
last equation in Eq. (2.18) leads to$^{[10,11]}$ 
\begin{equation}
\rho (\hat S_0)\hat \Gamma _n^d=0  \eqnum{2.25}
\end{equation}
where $\rho (\hat S_0)$ was defined in Eq. (2.19). In addition, when Eqs.
(2.13) and (2.12) are substituted into Eq. (2.15), one may obtain the ghost
equations satisfied by the finite part $\hat \Gamma _n^f$ and the divergent
part $\hat \Gamma _n^d$ of the functional $\hat \Gamma _n$. For the
divergent $\hat \Gamma _n^d$, we have$^{[10,11]}$ 
\begin{equation}
\frac{\delta \hat \Gamma _n^d}{\delta \bar C^a}-\partial _\mu (\frac{\delta 
\hat \Gamma _n^d}{\delta u_\mu ^a})=0.  \eqnum{2.26}
\end{equation}
It is emphasized here that even in the general gauges ($\alpha \neq 0)$, the
W-T identity and the ghost equation satisfied by the divergent functional $%
\hat \Gamma _n^d$ are the same as those given in the Landau gauge ($\alpha
=0 $) because the functions $\omega $ in Eq. (2.14) and the ${\mu }^2C^a(x)$
in Eq. (2.15) are finite.

Clearly, the divergent functional $\hat \Gamma _n^d$ and thus the
counterterm $\Delta \hat S_{n-1}$ may be determined by solving Eq. (2.25)
and (2.26) together, or, instead, by solving Eq. (2.21) and (2.26) provided
that the action $\hat S_{n-1}$ has been given in the former $n-1$ steps of
recursion. The general solution to the above equations was already found in
the literature$^{[10,11]}$. It consists of two parts as shown in the
following 
\begin{equation}
\Delta \hat S_{n-1}=-\hat \Gamma _n^d=\sum_\alpha a_\alpha ^nH_\alpha +\rho (%
\hat S_{n-1})F_n  \eqnum{2.27}
\end{equation}
where the $H_\alpha $ in the first term are functionals of the field
variables $A_\mu ^a,\bar \psi $ and $\psi $ which are invariant with respect
to the gauge transformation. Therefore, they obviously satisfy Eq. (2.21) or
(2.25). These functionals $H_\alpha $ can only come from the first four
terms in Eq. (2.4) and of the forms 
\begin{equation}
H_G=-\frac 14\int d^4x(\partial _\mu A_\nu ^a-\partial _\nu A_\mu
^a+gf^{abc}A_\mu ^bA_\nu ^c)^2,  \eqnum{2.28}
\end{equation}
\begin{equation}
H_F=\int d^4x\bar \psi i\gamma ^\mu (\partial _\mu -igA_\mu ^aT^a)\psi , 
\eqnum{2.29}
\end{equation}
\begin{equation}
H_M=\int d^4xM\bar \psi \psi  \eqnum{2.30}
\end{equation}
and 
\begin{equation}
H_m=\frac 12m^2\int d^4x(A_T^{a\mu })^2  \eqnum{2.31}
\end{equation}
here $H_m$ is written only for the transverse part of the vector potentials
because only for this part of the vector potential, the action of mass term
is gauge-invariant. The second term in Eq. (2.27) directly follows from the
nilpotency property of the operator $\rho $, $\rho ^2F_n=0$, with the
functional $F_n$ being arbitrary$^{[10]}$, as shown in Appendix.

It is noted that the term $\Delta \hat S_{n-1}$ as a part of the action
demands that the $F_n$ must be a functional with minus mass dimension and
minus ghost number so as to make the action to be dimensionless and of zero-
ghost number . This can be seen from the expression

\begin{equation}
\rho (\hat S_{n-1})F_n=\int d^4x\{\frac{\delta \hat S_{n-1}}{\delta \varphi
_i}\cdot \frac{\delta F_n}{\delta u_i}+\frac{\delta \hat S_{n-1}}{\delta u_i}%
\cdot \frac{\delta F_n}{\delta \varphi _i}\}.  \eqnum{2.32}
\end{equation}
by noticing that the ghost numbers of the functions $A_\mu ^a,\overline{\psi 
},\psi ,\overline{C}^a,C^a,u_\mu ^a,v^a,\overline{\zeta },\zeta $ are $%
0,0,0,-1,+1,-1,-2,-1,-1$ and the mass dimensions of these functions are $1,%
\frac 32,\frac 32,1,1,2,2,\frac 32,\frac 32$, respectively. Furthermore, the 
$F_n$, as easily seen, must satisfy the ghost equation. 
\begin{equation}
\frac{\delta F_n}{\delta \bar C^a}-\partial _\mu \frac{\delta F_n}{\delta
u_\mu ^a}=0.  \eqnum{2.33}
\end{equation}
With the requirements stated above, the form of the functional $F_n$ will be
uniquely determined, as given in the following 
\begin{equation}
F_n=\int d^4x\{b_A^nA^{a\mu }(u_\mu ^a-\partial _\mu \bar C^a)+\sum_{i\ne
A}b_i^n\varphi _iu_i\}.  \eqnum{2.34}
\end{equation}
It is noted that the coefficients $a_\alpha ^n$ in Eq. (2.27) and $b_i^n$ in
Eq. (2.34) all depend on the regularization parameter, say, the $\varepsilon
=2-n/2$ (which tends to zero, when $n\to 4$) in the dimensional
regularization$^{[21]}$. The operator $\rho (\hat S_{n-1})$ in Eq. (2.32)
implies that we have chosen the counterterm $\Delta \hat S_{n-1}$ to be the
solution of Eq. (2.21) for convenience of later recursion.

\section{Equivalence between the additional and multiplicative
renormalizations}

Up to the present, the counterterm $\Delta \hat S_{n-1}$ appearing in Eq.
(2.13) has explicitly been constructed as given in Eqs. (2.27)-(2.31) and
(2.34). The action $\hat S_{n-1}$ constructed in the foregoing steps of
recursion has the same functional structure as that for the $\hat S_0$ given
in Eq. (2.4). This can be seen from the fact that Eq. (2.21) has the same
form as Eq. (2.25). An interesting thing is that the role played by the
counterterm in Eq. (2.13) which has been found in Eq. (2.27) is only to make
a change to the variables of the first term $\hat S_{n-1}$ in Eq. (2.13). If
the coefficients in Eq. (2.34) are assumed to be infinitesimal, we have the
following variations: 
\begin{eqnarray}
\delta \varphi _i &=&\frac{\delta F_n}{\delta u_i},i=A,\bar \psi ,\psi ,C, 
\nonumber \\
\delta u_i &=&-\frac{\delta F_n}{\delta \varphi _i},i=\bar \psi ,\psi ,C, 
\nonumber \\
\delta u_\mu ^a-\partial _\mu \delta \bar C^a &=&-\frac{\delta F_n}{\delta
A_\mu ^a},i=A.  \eqnum{3.1}
\end{eqnarray}
According to the definition given in Eq. (2.24) and applying the ghost
equation obeyed by the action $\ \hat S_{n-1}$ 
\begin{equation}
\frac{\delta \ \hat S_{n-1}}{\delta \bar C^a}-\partial _\mu (\frac{\delta 
\hat S_{n-1}}{\delta u_\mu ^a})=0  \eqnum{3.2}
\end{equation}
and the variations defined in Eq. (3.1), one can derive$^{[10-12]}$%
\begin{equation}
\begin{array}{c}
\hat S_{n-1}[\varphi _i,u_i]+\rho (\hat S_{n-1})F_n[\varphi _i,u_i] \\ 
=\hat S_{n-1}[\varphi _i,u_i]+\frac{\delta \hat S_{n-1}}{\delta A_\mu ^a}%
\cdot \frac{\delta F_n}{\delta u^{a\mu }}+\frac{\delta \hat S_{n-1}}{\delta
u^{a\mu }}\cdot \frac{\delta F_n}{\delta A_\mu ^a} \\ 
+\sum\limits_{i\neq A}(\frac{\delta \hat S_{n-1}}{\delta \varphi _i}\cdot 
\frac{\delta F_n}{\delta u_i}+\frac{\delta \hat S_{n-1}}{\delta u_i}\cdot 
\frac{\delta F_n}{\delta \varphi _i}) \\ 
=\hat S_{n-1}[\varphi _i,u_i]+\sum\limits_{i=A,\psi ,\overline{\psi ,}%
C}(\delta \varphi _i\cdot \frac{\delta \hat S_{n-1}}{\delta \varphi _i}%
+\delta u_i\cdot \frac{\delta \hat S_{n-1}}{\delta u_i})+\delta \bar C%
^a\cdot \frac{\delta \hat S_{n-1}}{\delta \bar C^a} \\ 
=\hat S_{n-1}[\varphi _i^{\prime },u_i^{\prime }]
\end{array}
\eqnum{3.3}
\end{equation}
in which 
\begin{equation}
\begin{array}{c}
\varphi _i^{\prime }=\varphi _i+\delta \varphi _i=Y_i^n\varphi _i,\;\;\;i=A,%
\bar \psi ,\psi ,C, \\ 
u_i^{\prime }=u_i+\delta u_i=Y_i^{n^{-1}}u_i,\;\;\;i=A,\bar \psi ,\psi ,C,
\\ 
\bar C^{\prime }{}^a=\bar C^a+\delta \bar C^a=Y_A^{n^{-1}}\bar C^a
\end{array}
\eqnum{3.4}
\end{equation}
where the coefficients $Y_i^n$, according to the definitions in Eq. (3.1),
can be calculated from the expression written in Eq. (2.34). The results are 
\begin{equation}
Y_i^n=1+b_i^n,i=A,\bar \psi ,\psi ,C.  \eqnum{3.5}
\end{equation}
Considering that the functionals in the first term of Eq. (2.27) are
gauge-invariant and of the same functional structure as those terms in the $%
\hat S_{n-1}$ which are the functionals of the fields $A_\mu ^a$, $\bar \psi 
$ and $\psi $, we are allowed to change the field variables from $\varphi _i$
to $\varphi _i^{\prime }$ in the functionals in the first term of Eq. (2.27)
and combine these functionals with the corresponding terms in the $\hat S%
_{n-1}[\varphi _i^{\prime },u_i^{\prime }]$ shown in Eq. (3.3) together. The
results is just to redefine the variables and physical constants in the
action $\hat S_{n-1}$. Thus, the action in Eq. (2.13) can be written as 
\begin{equation}
\hat S_n[\varphi _i,u_i]=\hat S_{n-1}[\sqrt{Z_i^n}\varphi _i,\sqrt{\tilde Z%
_i^n}u_i]  \eqnum{3.6}
\end{equation}
where $Z_i^n$ and $\tilde Z_i^n$ are the $n$-th order multiplicative
renormalization constants for the field and source variables respectively.
Eq. (3.6) establishes a recursive relation of the renormalization. When the
order $n$ tends to infinity, we obtain from Eq. (3.6) by recursion the
following result 
\begin{equation}
\hat S[\varphi _i,u_i]=\hat S_0[\varphi _i^0,u_i^0]  \eqnum{3.7}
\end{equation}
where 
\begin{equation}
\begin{array}{c}
\varphi _i^0=\sqrt{Z_i}\varphi _i,\text{ }u_i^0=\sqrt{\tilde Z_i}u_i, \\ 
g^0=Z_gg,\text{ }M^0=Z_MM,\text{ }m^0=Z_mm
\end{array}
\eqnum{3.8}
\end{equation}
are the bare quantities appearing in the unrenormalized action $\hat S_0$.
The renormalization constants in Eq. (3.8) are defined by 
\begin{equation}
\begin{array}{c}
Z_i=\prod\limits_{n=1}^\infty Z_i^n,\text{ }\tilde Z_i=\prod\limits_{n=1}^%
\infty \tilde Z_i^n,\text{ }Z_g=\prod\limits_{n=1}^\infty Z_g^n, \\ 
Z_M=\prod\limits_{n=1}^\infty Z_M^n,\text{ }Z_m=\prod\limits_{n=1}^\infty
Z_m^n.
\end{array}
\eqnum{3.9}
\end{equation}
Eq. (3.7) shows us that the renormalized action has the same functional
structure as the unrenormalized one. In the derivation shown above, we start
from the additional renormalization by introducing the counterterms to
remove the divergences appearing in the functional $\hat \Gamma $ and then
obtain the result of multiplicative renormalization which is given by
introducing the renormalization constants to absorb the divergences and
redefining the field functions and physical parameters. This just shows the
equivalence between the both renormalizations.

To be more specific, let us describe the one-loop renormalization of the
functional $\Gamma _1$ starting from the action written in Eqs. (2.3) and
(2.4). For convenience of statement, we at first discuss the renormalization
in the Landau gauge in which the gluon fields are transverse. For the
one-loop renormalization, as indicated in Eqs. (2.9) and (2.10), we have to
introduce a counterterm whose general form was given in Eq. (2.27) with the
order label $n=1$. In the first term of Eq. (2.27), the gauge-invariant
functionals were written in Eqs .(2.28)-(2.31). The corresponding
coefficients in Eq.(2.27) will be written as $a_G,a_F,a_M$ and $a{_m}$. In
the following, the order label will be suppressed and the source terms in
Eq. (2.4) will be omitted for simplicity because these terms act only in the
intermediate stages of the proof. As demonstrated in Eq. (3.3), the
variables of the action $\hat S_0$ in Eq. (2.10) which was explicitly
written in Eq. (2.4) will be changed to the ones shown in Eq.(3.4) owing to
the effect of the counterterm given by the second term in Eq. (2.27) with
the functional $F$ being represented in Eq. (2.34) and the variables in the
counterterms of the first term in Eq. (2.27) which are explicitly written in
Eqs. (2.28)-(2.31) may also be made such changes due to the gauge-invariance
of the functionals. Thus, the action in Eq. (2.10) without the source terms
may be written as 
\begin{equation}
\begin{array}{c}
\hat S[A_\mu ^a,\bar \psi ,\psi ,\bar C^a,C^a] \\ 
\int d^4x\{Y_FY_{\overline{\psi }}Y_\psi \bar \psi i\gamma ^\mu (\partial
_\mu -igY_AA_\mu ^aT^a)\psi -MY_MY_{\overline{\psi }}Y_\psi \bar \psi \psi
\\ 
-\frac 14Y_A^2Y_G(\partial _\mu A_\nu ^a-\partial _\nu A_\mu
^a+gY_Af^{abc}A_\mu ^bA_\nu ^c)^2+\frac 12Y_mY_A^2m^2A^{a\mu }A_\mu ^a \\ 
+Y_A^{-1}Y_C\bar C^a\Box C^a+Y_Cgf^{abc}\partial ^\mu \bar C^aC^bA_\mu ^c\}
\end{array}
\eqnum{3.10}
\end{equation}
where 
\begin{equation}
Y_F=1+a_F,Y_M=1+a_M,Y_G=1+a_G,Y_m=1+a_m  \eqnum{3.11}
\end{equation}
and the subscript $T$ marking the transverse gauge fields $A_T^{a\mu }$ has
been suppressed for simplicity of representation. When we define
renormalization constants by$^{[11-12]}$%
\begin{equation}
\begin{array}{c}
Z_2=Y_FY_{\overline{\psi }}Y_\psi ,\text{ }Z_3=Y_GY_A^2,\text{ }\tilde Z%
_3=Y_CY_A^{-1}, \\ 
Z_1^F=Y_FY_{\overline{\psi }}Y_\psi Y_A,\text{ }Z_1=Y_GY_A^3,\text{ }%
Z_4=Y_GY_A^4, \\ 
\tilde Z_1=Y_C,\text{ }Z_M=Y_MY_F^{-1},\text{ }%
Z_m^2=Y_mY_A^2Z_3^{-1}=Y_mY_G^{-1}
\end{array}
\eqnum{3.12}
\end{equation}
and noticing the relation given in Eq.(2.3), We may write the full action as
follows 
\begin{equation}
\begin{array}{c}
S=\int d^4x\{Z_2\bar \psi (i\gamma ^\mu \partial _\mu -MZ_M)\psi +gZ_1^F\bar 
\psi \gamma ^\mu A_\mu ^aT^a\psi \\ 
-\frac 14Z_3(\partial _\mu A_\nu ^a-\partial _\nu A_\mu ^a)^2-\frac 12%
gZ_1f^{abc}(\partial _\mu A_\nu ^a-\partial _\nu A_\mu ^a)A^{b\mu }A^{c\nu }
\\ 
-\frac 14g^2Z_4f^{abc}f^{ade}A^{b\mu }A^{c\nu }A_\mu ^dA_\nu ^e+\frac 12%
Z_m^2Z_3m^2A_\mu ^aA^{a\mu } \\ 
+\tilde Z_3\bar C^a\Box C^a+g\tilde Z_1f^{abc}\partial ^\mu \bar C^aC^bA_\mu
^c-\frac 1{2\alpha }(\partial ^\mu A_\mu ^a)^2\}.
\end{array}
\eqnum{3.13}
\end{equation}
It is noted here that the last equality in Eq. (3.12) gives the definition
of gluon mass renormalization constant as $Z_m=Z_3^{-1/2}Y_m^{1/2}Y_A$ which
is consistent with that given in the renormalization of gluon propagator as
shown in Eq. (4.25) of paper I. The action in Eq. (3.13) with the
renormalization constants being defined in Eq. (3.12) just gives the
recursive relation shown in Eq. (3.6) with the label $n=1$. This action
would eliminate the divergence appearing in the generating functional $%
\Gamma $ evaluated in the one-loop approximation.

From Eq. (3.12), it is clear to see that 
\begin{equation}
\frac{Z_1}{Z_3}=\frac{Z_1^F}{Z_2}=\frac{\tilde Z_1}{\tilde Z_3}=\frac{Z_4}{%
Z_1}.  \eqnum{3.14}
\end{equation}
This is the Slavnov-Taylor (S-T) identity satisfied by the renormalization
constants which is completely the same as given in the massless QCD$%
^{[16,17]}$.

If we define the bare quantities as$^{[11,12]}$%
\begin{equation}
\begin{array}{c}
A_0^{a\mu }=\sqrt{Z_3}A^{a\mu },\;\psi _0=\sqrt{Z_2}\psi ,\;\bar \psi _0=%
\sqrt{Z_2}\bar \psi , \\ 
C_0^a=\sqrt{\tilde Z_3}C^a,\;\bar C_0^a=\sqrt{\tilde Z_3}\bar C%
^a,\;g_0=Z_1Z_3^{-3/2}g, \\ 
M_0=Z_MM,\;m_0=Z_mm,\;\alpha _0=Z_3\alpha
\end{array}
\eqnum{3.15}
\end{equation}
and use the identity in Eq. (3.14), we arrive at 
\begin{equation}
S[A_\mu ^a,\bar \psi ,\psi ,\bar C^a,C^a]=S_0[A_0^{a\mu },\bar \psi _0,\psi
_0,\bar C_0^a,C_0^a]  \eqnum{3.16}
\end{equation}
where 
\begin{equation}
\begin{array}{c}
S_0=\int d^4x\{\bar \psi _0[i\gamma _\mu (\partial ^\mu -ig_0A_0^{a\mu
}T^a)-M_0]\psi _0 \\ 
-\frac 14(\partial ^\mu A_0^{a\nu }-\partial ^\nu A_0^{a\mu
}+g_0f^{abc}A_0^{b\mu }A_0^{c\nu })^2 \\ 
+\bar C_0^a\Box C_0^a+g_0f^{abc}\partial _\mu \bar C_0^aC_0^bA_0^{c\mu } \\ 
+\frac 12m_0^2A_0^{a\mu }A_{0\mu }^a-\frac 1{2\alpha _0}(\partial _\mu
A_0^{a\mu })^2\}
\end{array}
\eqnum{3.17}
\end{equation}
is the unrenormalized action. Eqs. (2.3), (2.4), (3.16) and (3.17) indicate
that the actions, renormalized and unrenormalized, have the same structure
and thus the same symmetry, just as we met in the massless gauge field
theory.

We note here that although the above results are obtained in the one-loop
renormalization, they can, actually, be considered to be the exact ones. In
fact, for removing the two-loop divergence in the $\hat \Gamma $, obviously,
the second cycle of recursion of the renormalization can be carried out in
the same way as stated in Eqs. (3.10)-(3.13) by starting from the action
shown in Eq. (3.10) and higher order recursions can be continued further
along the same line. The results given in Eqs. (3.10)-(3.17) formally remain
unchanged for each cycle of the recursion. Therefore, by the recursive
procedure, all the results denoted in Eqs.(3.12)-(3.17) can be regarded as
the ones as shown in Eqs. (3.7)-(3.9).

Now, let us describe the renormalizability of the theory in the general
gauges. We firstly note that the results in the Landau gauge as given before
can readily be extended to the other gauges. As mentioned before, in the
general gauges, we still have the equations written in Eqs. (2.21) and
(2.26) for the divergent part $\hat \Gamma _n^d$. Therefore, the
counterterm, as the solution to the equations (2.21) and (2.26), is still
expressed by Eqs. (2.27)-(2.31) and (2.34) with a note that except for the
gluon mass counterterm written in Eq. (2.31) which is still given by the
transverse field, the vector potential in the other counterterms now becomes
the full one. It is clear that the statements in Eqs. (3.1)-(3.9) completely
hold for the general gauges. The results described in Eqs. (3.10)-(3.17), as
easily seen, except for a few supplements for the gluon and ghost particle
mass terms, are also preserved in the present case. The gluon mass term in
Eq. (2.4) is now written for the full vector potential. When the countertems
in Eq. (2.27) with their explicit expressions denoted in Eqs. (2.28)-(2.31)
and (2.34) are added to the action, the mass term becomes 
\begin{equation}
\int d^4x\frac 12Y_mY_A^2m^2[A_T^{a\mu }A_{T\mu }^a+Y_m^{-1}A_L^{a\mu
}A_{L\mu }^a].  \eqnum{3.18}
\end{equation}
This term should replace the corresponding term in Eq. (3.10) to appear in
the action. In the above, the factor $Y_A^2$ arises from the variable change
generated by the counterterm given in the second term in Eq. (2.27) as
described in Eqs .(3.3) and (3.4) and the factor $Y_m$ comes from the
counterterm denoted in Eq. (2.31) which is written only for the transverse
fields . When we notice the last equality in Eq. (3.12) and define 
\begin{equation}
Z_3^{\prime }=Y_m^{-1}  \eqnum{3.19}
\end{equation}
and

\begin{equation}
A^{a\mu }=A_T^{a\mu }+\sqrt{Z_3^{\prime }}A_L^{a\mu }.  \eqnum{3.20}
\end{equation}
Eq. (3.18) can be written as

\begin{equation}
\frac 12\int d^4xZ_m^2Z_3m^2(A_T^{a\mu }A_{T\mu }^a+Z_3^{\prime }A_L^{a\mu
}A_{L\mu }^a)=\frac 12\int d^4xZ_m^2Z_3m^2A^{a\mu }A_\mu ^a  \eqnum{3.21}
\end{equation}
where the orthonormality between the transverse and longitudinal vector
potentials has been considered. With the above expression, the gluon mass
term in Eq. (3.13) can be understood for the full vector potential in the
general gauge. Eq. (3.21) shows that the renormalization of the longitudinal
part of the vector potential is different from that for the transverse part
by an extra renormalization constant $\sqrt{Z_3^{\prime }}$. This result is
consistent with the renormalization of gluon propagator shown in section 4
of paper I where the longitudinal part of the propagator has an extra
renormalization constant $Z_3^{\prime }$. According to the definition given
in Eq. (3.15), the gluon mass term in Eq. (3.21) will come to the form
expressed by the bare quantities as shown in Eq. (3.17).

Let us turn to the renormalization of the ghost particle mass term. As
mentioned in section 2, the ghost particle mass term $\mu ^2C^a$ in Eq.
(2.15) which appears in general gauges is finite and therefore it is not
included in the ghost equation (2.26) satisfied by the counterterm $\hat 
\Gamma _n^d$. That is to say, the ghost particle term $\bar C^a{\mu }^2C^a$
in the Lagrangian remains unchanged in the process of divergence subtraction
because it is not affected by the counterterm. The counterterm only changes
the kinetic energy term $\bar C^a\Box C^a$ and the interaction term $%
gf^{abc}\partial ^\mu \bar C^aC^bA_\mu ^c$ as for the case in the Landau
gauge. This fact coincides with the renormalization of ghost particle
propagator described in section 4 of paper I where the ghost particle
self-energy is written as $\Omega (k)=k^2\hat \Omega (k^2)$ and combined
with $k^2$ which comes from the kinetic energy operator $\Box $. Therefore,
in the general gauge, the ghost particle kinetic energy term in Eq. (3.10)
may be extended to the following form by directly adding the ghost particle
mass term 
\begin{equation}
Y_A^{-1}Y_C\bar C^a\Box C^a+\bar C^a{\mu }^2C^a  \eqnum{3.22}
\end{equation}
where the factors ${Y_A^{-1}}$ and ${Y_C}$ arise respectively from the
change of the variables $\bar C^a$ and ${C}^a$ which are caused by the
counterterm contained in the second term in Eq. (2.27), as described in Eqs.
(3.3) and (3.4). With the definitions given in Eqs. (3.12) and (3.15), Eq.
(3.22) can be represented as 
\begin{equation}
\tilde Z_3\bar C^a(\Box +\tilde Z_3^{-1}{\mu }^2)C^a=\bar C_0^a(\Box +{\mu }%
_0^2)C_0^a  \eqnum{3.23}
\end{equation}
where 
\begin{equation}
\mu =\sqrt{\tilde Z_3}\mu _0  \eqnum{3.24}
\end{equation}
which just is the relation given by Eq. (4.30) in paper I. Clearly, in the
general gauge, the ghost particle kinetic energy terms in Eqs. (3.13) and
(3.17) should be replaced by the terms on the LHS and RHS of Eq. (3.23),
respectively.

\section{Gluon three-line vertex}

~In the preceding section, the S-T identity shown in Eq. (3.14) for the
renormalization constants was derived from the general proof of the
renormalizability of the QCD with massive gluons. Originally, the same
identity given in the QCD\ with massless gluons was derived from the W-T
identities satisfied by the proper vertices and their renormalization$%
^{[16,17]}$. Following the similar line, we will re-derive the identity in
Eq. (3.14) so as to give a check of the identity and its derivation. Since
the difference between the massive QCD and the massless QCD mainly lies in
the gluon field, in this section and the next section, we limit ourself to
take the gluon three-line and four-line proper vertices as examples to show
how to derive W-T identities satisfied by the vertices and discuss their
renormalization. To derive W-T identities for the vertices, it is suitable
to employ the identities written in Eqs. (2.14) and (2.15) though, we would
like to start from the identities satisfied by the generating functional of
Green functions in order to see, by the way, the difference between the
Green functions for the massive gluons and the massless ones. Let us begin
with derivation of an identity satisfied by the gluon three-point Green
function from the following W-T identity 
\begin{eqnarray}
\frac 1\alpha \partial _x^\mu \frac{\delta Z[J]}{\delta J^{a\mu }(x)}+\int
d^4yJ^{b\nu }(y)\frac{\delta ^2Z[J,K,u]}{\delta K^a(x)\delta u^{b\nu }(y)}%
|_{K=u=0}=0  \eqnum{4.1}
\end{eqnarray}
and the ghost equation 
\begin{equation}
\begin{array}{c}
i\partial _\mu ^x\frac{\delta ^2Z[J.K.u]}{\delta u_\mu ^a(x)\delta K^b(y)}%
|_{K=u=0}+i{\mu }^2\frac{\delta ^2Z[J,\bar K,K]}{\delta \bar K^a(x)\delta
K^b(y)}|_{\bar K=K=0} \\ 
+\delta ^{ab}\delta ^4(x-y)Z[J]=0
\end{array}
\eqnum{4.2}
\end{equation}
which were derived in section 4 of paper I ( Hereafter, $\mu $ denotes the
unrenormalized mass of ghost particle and $\mu _R$ designates the
renormalized mass of the particle). By taking successive differentiations of
Eq. (4.1) with respect to the sources $J_\nu ^b(y)$ and $J_\lambda ^c(z)$
and then setting the sources to vanish, one may obtain the W-T identity
obeyed by the gluon three-point Green function which is written in the
operator form as 
\begin{eqnarray}
\frac 1\alpha \partial _x^\mu G_{\mu \nu \lambda }^{abc}(x,y,z)
&=&<0^{+}|T^{*}[\hat {\bar C^a}(x)\hat D_\nu ^{bd}(y)\hat C^d(y)\hat A%
_\lambda ^c(z)]|0^{-}>  \nonumber \\
+ &<&0^{+}|T^{*}[\hat {\bar C^a}(x)\hat A_\nu ^b(y)\hat D_\lambda ^{cd}(z)%
\hat C^d(z)]|0^{-}>  \eqnum{4.3}
\end{eqnarray}
where 
\begin{equation}
G_{\mu \nu \lambda }^{abc}(x,y,z)=<0^{+}|T[\hat A_\mu ^a(x)\hat A_\nu ^b(y)%
\hat A_\lambda ^c(z)]|0^{-}>  \eqnum{4.4}
\end{equation}
is the three-point Green function mentioned above. The identity in Eq. (4.3)
will be simplified by a ghost equation which may be derived by
differentiating Eq. (4.2) with respect to the source $J_\lambda ^c(z)$ 
\begin{equation}
\begin{array}{c}
\partial _x^\mu <0^{+}|T^{*}\{\hat D_\mu ^{ad}(x)\hat C^d(x)\hat {\bar C^b}%
(y)\hat A_\lambda ^c(z)\}|0^{-}> \\ 
+{\mu }^2<0^{+}|T[\hat C^a(x)\hat {\bar C^b}(y)\hat A_\lambda
^c(z)]|0^{-}]>=0.
\end{array}
\eqnum{4.5}
\end{equation}
Taking derivatives of Eq. (4.3) with respect to $y$ and $z$ and employing
Eq. (4.5), we get 
\begin{equation}
\partial _x^\mu \partial _y^\nu \partial _z^\lambda G_{\mu \nu \lambda
}^{abc}(x,y,z)=\alpha \mu ^2\{\partial _y^\nu G_{~~\nu
}^{cab}(z,x,y)+\partial _z^\lambda G_{~~\lambda }^{bac}(y,x,z)\}  \eqnum{4.6}
\end{equation}
where 
\begin{equation}
G_{~~\mu }^{abc}(x,y,z)=<0^{+}|T\{\hat C^a(x)\hat {\bar C^b}(y)\hat A_\mu
^c(z)\}|0^{-}>.  \eqnum{4.7}
\end{equation}
In the Landau gauge or the zero-mass limit ($\mu =0$), Eq. (4.6) reduces to 
\begin{equation}
\partial _x^\mu \partial _y^\nu \partial _z^\lambda G_{\mu \nu \lambda
}^{abc}(x,y,z)=0  \eqnum{4.8}
\end{equation}
which shows the transversity of the Green function. From Eq. (4.6), we may
derive a W-T identity for the gluon three-line vertex. For this purpose, it
is necessary to use the following one-particle-irreducible decompositions of
the Green functions$^{[11,12]}$%
\begin{equation}
\begin{array}{c}
G_{\mu \nu \lambda }^{abc}(x,y,z)=\int d^4x^{\prime }d^4y^{\prime
}d^4z^{\prime }iD_{\mu \mu ^{\prime }}^{aa^{\prime }}(x-x^{\prime }) \\ 
\times iD_{\nu \nu ^{\prime }}^{bb^{\prime }}(y-y^{\prime })iD_{\lambda
\lambda ^{\prime }}^{cc^{\prime }}(z-z^{\prime })\Gamma _{a^{\prime
}b^{\prime }c^{\prime }}^{\mu ^{\prime }\nu ^{\prime }\lambda ^{\prime
}}(x^{\prime },y^{\prime },z^{\prime },)
\end{array}
\eqnum{4.9}
\end{equation}
and 
\begin{equation}
\begin{array}{c}
G_{~~\,\nu }^{abc}(x,y,z)=\int d^4x^{\prime }d^4y^{\prime }d^4z^{\prime
}i\Delta ^{aa^{\prime }}(x-x^{\prime })\Gamma ^{a^{\prime }b^{\prime
}c^{\prime },\nu ^{\prime }}(x^{\prime },y^{\prime },z^{\prime }) \\ 
\times i\Delta ^{b^{\prime }b}(y^{\prime }-y)iD_{\nu ^{\prime }\nu
}^{c^{\prime }c}(z^{\prime }-z)
\end{array}
\eqnum{4.10}
\end{equation}
where $iD_{\mu \mu ^{\prime }}^{aa^{\prime }}(x-x^{\prime })$ and $i\Delta
^{aa^{\prime }}(x-x^{\prime })$ are the propagators for the gluon and the
ghost particle respectively which were derived in paper I, $\Gamma
_{abc}^{\mu \nu \lambda }(x,y,z)$ and $\Gamma _{~~\lambda }^{abc}(x,y,z)$
are the three-line gluon proper vertex and the three-line ghost-gluon proper
vertex respectively. They are defined as$^{[11,12]}$%
\begin{equation}
\Gamma _{abc}^{\mu \nu \lambda }(x,y,z)=i\frac{\delta ^3\Gamma }{\delta
A_\mu ^a(x)\delta A_\nu ^b(y)\delta A_\lambda ^c(z)}|_{J=0}  \eqnum{4.11}
\end{equation}
and 
\begin{equation}
\Gamma _{~~\lambda }^{abc}(x,y,z)=\frac{\delta ^3\Gamma }{i\delta \bar C%
^a(x)\delta C^b(y)\delta A^{c\lambda }(z)}|_{J=0}  \eqnum{4.12}
\end{equation}
here$J$ stands for all the external sources. Substituting Eqs. (4.9) and
(4.10) into Eq. (4.6) and transforming Eq. (4.6) into the momentum space,
one can derive an identity which establishes the relation between the
longitudinal part of three-line gluon vertex and the three-line ghost-gluon
vertex as follows 
\begin{equation}
\begin{array}{c}
p^\mu q^\nu k^\lambda \Lambda _{\mu \nu \lambda }^{abc}(p,q,k)=-\frac{\mu ^2}%
\alpha \chi (p^2)[\chi (k^2)q^\nu \Lambda _{~~\nu }^{cab}(k,p,q) \\ 
+\chi (q^2)k^\lambda \Lambda _{~~\lambda }^{bac}(q,p,k)]
\end{array}
\eqnum{4.13}
\end{equation}
where we have defined 
\begin{equation}
\begin{array}{c}
\Gamma _{\mu \nu \lambda }^{abc}(p,q,k)=(2\pi )^4\delta ^4(p+q+k)\Lambda
_{\mu \nu \lambda }^{abc}(p,q,k) \\ 
\Gamma _{~~\lambda }^{abc}(p,q,k)=(2\pi )^4\delta ^4(p+q+k)\Lambda
_{~~\lambda }^{abc}(p,q,k)
\end{array}
\eqnum{4.14}
\end{equation}
and 
\begin{equation}
\chi (p^2)=\{k^2[1+\hat \Pi _L(p^2)]-\mu ^2+i\varepsilon \}\{k^2[1+\hat 
\Omega (p^2)]-\mu ^2+i\varepsilon \}^{-1}  \eqnum{4.15}
\end{equation}
here $\hat \Pi _L(p^2)$ and $\hat \Omega (p^2)$ are the self-energies
appearing in the longitudinal part of gluon propagator and the ghost
particle propagator (see Eqs. (4.11) and (4.22) in paper I).

Obviously, in the Landau gauge, Eq. (4.13) reduces to 
\begin{equation}
p^\mu q^\nu k^\lambda \Lambda _{\mu \nu \lambda }^{abc}(p,q,k)=0 
\eqnum{4.16}
\end{equation}
which implies that the vertex is transverse in this case. In the lowest
order approximation, owing to 
\begin{equation}
\chi (p^2)=1  \eqnum{4.17}
\end{equation}
and 
\begin{equation}
\Lambda _{~~~~~\mu }^{(0)abc}(p,q,k)=gf^{abc}p_\mu ,  \eqnum{4.18}
\end{equation}
the RHS of Eq. (4.13) vanishes, therefore, we have 
\begin{equation}
p^\mu q^\nu k^\lambda \Lambda _{~~~\mu \nu \lambda }^{(0)abc}(p,q,k)=0. 
\eqnum{4.19}
\end{equation}
This result is consistent with that for the bare three-line gluon vertex
given by the Feynman rule.

Now, let us discuss renormalization of the three-line gluon vertex. At
first, we note that the notation used here is different from the previous
section. In this section, a renornalized quantity will be marked with a
subscript $R$; a quantity without the subscript $R$ represents the
unrenormalized one. From Eq. (3.15) or the renormalization of the gluon and
ghost particle propagators as described in paper I, we see 
\begin{equation}
\begin{array}{c}
A^{a\mu }(x)=\sqrt{Z_3}A_R^{a\mu }(x), \\ 
C^a(x)=\sqrt{\tilde Z_3}C_R^a(x),\text{ }\bar C^a(x)=\sqrt{\tilde Z_3}\bar C%
_R^a(x).
\end{array}
\eqnum{4.20}
\end{equation}
According to these relations and the definitions given in Eqs. (4.11),
(4.12) and (4.14), we find 
\begin{equation}
\begin{array}{c}
\Lambda _{\mu \nu \lambda }^{abc}(p,q,k)=Z_3^{-3/2}\Lambda _{R\mu \nu
\lambda }^{~~abc}(p,q,k), \\ 
\Lambda _{~~\lambda }^{abc}(p,q,k)=\tilde Z_3^{-1}Z_3^{-1/2}\Lambda
_{R~~\lambda }^{~~abc}(p,q,k).
\end{array}
\eqnum{4.21}
\end{equation}
Applying these relations, the renormalized version of the identity written
in Eq. (4.13) will be 
\begin{equation}
\begin{array}{c}
p^\mu q^\nu k^\lambda \Lambda _{R\mu \nu \lambda }^{~~abc}(p,q,k)=-\frac{\mu
_R^2}{\alpha _R}{\chi }_R(p^2)[{\chi _R}(k^2)q^\nu \Lambda _{R~~\nu
}^{~~cab}(k,p,q) \\ 
+\chi _R(q^2)k^\lambda \Lambda _{R~~\lambda }^{~~bac}(q,p,k)]
\end{array}
\eqnum{4.22}
\end{equation}
where $\alpha _R=Z_3^{-1}\alpha $, $\mu _R=\sqrt{\tilde Z_3}\mu $ and 
\begin{equation}
\chi _R(k^2)=\frac 1{1+\Omega _R(k^2)]}  \eqnum{4.23}
\end{equation}
is the renormalized expression of the function $\chi (k^2)$ which may be
obtained by substituting the relation (see Eq. (4.23) in paper I) 
\begin{equation}
\hat \Pi _L(k^2)=\frac{\mu ^2\hat \Omega (k^2)}{k^2[1+\hat \Omega (k^2)]} 
\eqnum{4.24}
\end{equation}
into Eq. (4.15) and then using the definitions 
\begin{equation}
\hat \Omega (k^2)=\hat \Omega (\nu ^2)+\hat \Omega ^c(k^2),\text{ }\tilde Z%
_3^{-1}=1+\hat \Omega (\nu ^2),\;\Omega _R(k^2)=\tilde Z_3\hat \Omega ^c(k^2)
\eqnum{4.25}
\end{equation}
which were ever shown in section 4 of paper I. At the renormalization point
chosen to be $p^2=q^2=k^2=\mu _R^2$, we see, $\chi _R(\mu _R^2)=1$. In this
case, the renormalized ghost-gluon vertex takes the form of the bare vertex
so that the RHS of Eq. (4.22) vanishes, therefore, we have 
\begin{equation}
p^\mu q^\nu k^\lambda \Lambda _{R\mu \nu \lambda
}^{~~abc}(p,q,k)|_{p^2=q^2=k^2=\mu _R^2}=0.  \eqnum{4.26}
\end{equation}

Ordinarily, one is interested in discussing the renormalization of such
three-line vertices that they are defined from the vertices defined in Eqs.
(4.11) and (4.12) by extracting a coupling constant $g$. These vertices are
denoted by $\tilde \Lambda _{\mu \nu \lambda }^{abc}(p,q,k)$ and $\tilde 
\Lambda _{~~\lambda }^{abc}(p,q,k)$. Commonly, they are renormalized in such
a fashion$^{[11,12]}$. 
\begin{equation}
\begin{array}{c}
\tilde \Lambda _{\mu \nu \lambda }^{abc}(p,q,k)=Z_1^{-1}\widetilde{\Lambda }%
_{R\mu \nu \lambda }^{abc}(p,q,k), \\ 
\tilde \Lambda _{~~\lambda }^{abc}(p,q,k)=\widetilde{Z}_1^{-1}\widetilde{%
\Lambda }_{R~~\lambda }^{abc}(p,q,k)
\end{array}
\eqnum{4.27}
\end{equation}
where $Z_1$ and $\tilde Z_1$ are referred to as the renormalization
constants for the three-line gluon vertex and the ghost-gluon vertex,
respectively. It is clear that the W-T identity shown in Eq. (4.13) also
holds for the vertices $\tilde \Lambda _{\mu \nu \lambda }^{abc}(p,q,k)$ and 
$\tilde \Lambda _{~~\lambda }^{abc}(p,q,k)$. So, when the vertices $\Lambda
_{\mu \nu \lambda }^{abc}(p,q,k)$ and $\Lambda _{~~\lambda }^{abc}(p,q,k)$
in Eqs. (4.13) are replaced by $\tilde \Lambda _{\mu \nu \lambda
}^{abc}(p,q,k)$ and $\tilde \Lambda _{~~\lambda }^{abc}(p,q,k)$ respectively
and then Eq. (4.27) is inserted to such an identity, we obtain a
renormalized version of the identity as follows 
\begin{equation}
\begin{array}{c}
p^\mu q^\nu k^\lambda \tilde \Lambda _{R\mu \nu \lambda }^{~~abc}(p,q,k)=-%
\frac{Z_1\tilde Z_3}{Z_3\tilde Z_1}\frac{\mu _R^2}{\alpha _R}\chi
_R(p^2)[\chi _R(k^2) \\ 
\times q^\nu \tilde \Lambda _{R~~\nu }^{~~cab}(k,p,q)+\chi _R(q^2)k^\lambda 
\tilde \Lambda _{R~~\lambda }^{~~bac}(q,p,k)].
\end{array}
\eqnum{4.28}
\end{equation}
When multiplying the both sides of Eq. (4.28) with a renormalized coupling
constant $g_R$ and absorbing it in the vertices, noticing 
\begin{equation}
\begin{array}{c}
\Lambda _{R\mu \nu \lambda }^{abc}(p,q,k)=g_R\widetilde{\Lambda }_{R\mu \nu
\lambda }^{abc}(p,q,k), \\ 
\Lambda _{R~~\lambda }^{abc}(p,q,k)=g_R\widetilde{\Lambda }_{R~~\lambda
}^{abc}(p,q,k),
\end{array}
\eqnum{4.29}
\end{equation}
we have 
\begin{equation}
\begin{array}{c}
p^\mu q^\nu k^\lambda \Lambda _{R\mu \nu \lambda }^{~~abc}(p,q,k)=-\frac{Z_1%
\tilde Z_3}{Z_3\tilde Z_1}\frac{\mu _R^2}{\alpha _R}\chi _R(p^2)[\chi _R(k^2)
\\ 
\times q^\nu \Lambda _{R~~\nu }^{~~cab}(k,p,q)+\chi _R(q^2)k^\lambda \Lambda
_{R~~\lambda }^{~~bac}(q,p,k)].
\end{array}
\eqnum{4.30}
\end{equation}
In comparison of Eq. (4.30) with Eq. (4.22), we see, except for the factor $%
Z_1\tilde Z_3Z_3^{-1}\tilde Z_1^{-1}$, the both identities are identical to
each other. From this observation, we deduce 
\begin{equation}
\frac{Z_1}{Z_3}=\frac{\tilde Z_1}{\tilde Z_3}.  \eqnum{4.31}
\end{equation}
This is the S-T identity which coincides with the one given in Eq. (3.14).

\section{Gluon four-line vertex}

By the similar procedure as deriving Eqs. (4.3) and (4.4), the W-T identity
obeyed by the gluon four-point Green function may be derived by
differentiating Eq. (4.1) with respect to the sources $J_\mu ^b(y),J_\lambda
^c(z)$ and $J_\tau ^d(u)$. The result is as follows$^5$%
\begin{equation}
\begin{array}{c}
\frac 1\alpha \partial _x^\mu G_{\mu \nu \lambda \tau }^{abcd}(x,y,z,u) \\ 
=<0^{+}|T^{*}[\hat {\bar C^a}(x)\hat D_\nu ^{be}(y)\hat C^e(y)\hat A_\lambda
^c(z)\hat A_\tau ^d(u)]|0^{-}> \\ 
+<0^{+}|T^{*}[\hat {\bar C^a}(x)\hat A_\nu ^b(y)\hat D_\lambda ^{ce}(z)\hat C%
^e(z)\hat A_\tau ^d(u)]|0^{-}> \\ 
+<0^{+}|T^{*}[\hat {\bar C^a}(x)\hat A_\nu ^b(y)\hat A_\lambda ^c(z)\hat D%
_\tau ^{de}(u)\hat C^e(u)]|0^{-}>
\end{array}
\eqnum{5.1}
\end{equation}
where 
\begin{equation}
G_{\mu \nu \lambda \tau }^{abcd}(x,y,z,u)=<0^{+}|T[\hat A_\mu ^a(x)\hat A%
_\nu ^b(y)\hat A_\lambda ^c(z)\hat A_\tau ^d(u)]|0^{-}>  \eqnum{5.2}
\end{equation}
is the gluon four-point Green function. The accompanying ghost equation may
be obtained by differentiating Eq. (4.2) with respect to the sources $%
J_\lambda ^c(z)$ and $J_\tau ^d(u)$. The result is 
\begin{equation}
\begin{array}{c}
\partial _x^\mu <0^{+}|T^{*}[\hat D_\mu ^{ae}(x)\hat C^e(x)\hat {\bar C^b}(y)%
\hat A_\lambda ^c(z)\hat A_\tau ^d(u)]|0^{-}> \\ 
+{\mu }^2G_{~~\lambda \tau }^{abcd}(x,y,z,u)=-\delta ^{ab}\delta
^4(x-y)D_{\lambda \tau }^{cd}(z-u)
\end{array}
\eqnum{5.3}
\end{equation}
where 
\begin{equation}
G_{~~\lambda \tau }^{abcd}(x,y,z,u)=<0^{+}|T[\hat C^a(x)\hat {\bar C^b}(y)%
\hat A_\lambda ^c(z)\hat A_\tau ^d(u)]|0^{-}>  \eqnum{5.4}
\end{equation}
is the four-point gluon-ghost particle Green function. Differentiation of
Eq. (5.1) with respect to the coordinates $y$, $z$ and $u$ and use of Eq.
(5.3) lead to 
\begin{equation}
\begin{array}{c}
\frac 1\alpha \partial _x^\mu \partial _y^\nu \partial _z^\lambda \partial
_u^\tau G_{\mu \nu \lambda \tau }^{abcd}(x,y,z,u)=\delta ^{ab}\delta
^4(x-y)\partial _z^\lambda \partial _u^\tau D_{\lambda \tau }^{cd}(z-u) \\ 
+\delta ^{ac}\delta ^4(x-z)\partial _y^\nu \partial _u^\tau D_{\nu \tau
}^{bd}(y-u)+\delta ^{ad}\delta ^4(x-u)\partial _y^\nu \partial _z^\lambda
D_{\nu \lambda }^{bc}(y-z) \\ 
+\mu ^2\{\partial _z^\lambda \partial _u^\tau G_{~~\lambda \tau
}^{bacd}(y,x,z,u)+\partial _y^\nu \partial _u^\tau G_{~~\nu \tau
}^{cabd}(z,x,y,u) \\ 
+\partial _y^\nu \partial _z^\lambda G_{~~\nu \lambda }^{dabc}(u,x,y,z)\}.
\end{array}
\eqnum{5.5}
\end{equation}

It is noted that the four-point Green functions appearing in the above
equations are unconnected. Their decompositions to connected Green functions
are not difficult to be found by making use of the relation between the
generating functionals $Z$ for the full Green functions and $W$ for the
connected Green functions, $Z=\exp (iW)$. The result is 
\begin{equation}
\begin{array}{c}
G_{\mu \nu \lambda \tau }^{abcd}(x,y,z,u)=G_{\mu \nu \lambda \tau
}^{abcd}(x,y,z,u)_c-D_{\mu \nu }^{ab}(x-y)D_{\lambda \tau }^{cd}(z-u) \\ 
-D_{\mu \lambda }^{ac}(x-z)D_{\nu \tau }^{bd}(y-u)-D_{\mu \tau
}^{ad}(x-u)D_{\nu \lambda }^{bc}(y-z)
\end{array}
\eqnum{5.6}
\end{equation}
and 
\begin{equation}
G_{~~\lambda \tau }^{abcd}(x,y,z,u)=G_{~~\lambda \tau
}^{abcd}(x,y,z,u)_c-\Delta ^{ab}(x-y)D_{\lambda \tau }^{cd}(z-u). 
\eqnum{5.7}
\end{equation}
The first terms marked by the subscript ''$c$'' in Eqs. ( 5.6) and (5.7) are
connected Green functions. When inserting Eqs. (5.6) and (5.7) into Eq.
(5.5) and using the W-T identity satisfied by the gluon propagator which was
derived in section 4 of paper I 
\begin{equation}
\partial _x^\mu \partial _y^\nu D_{\mu \nu }^{ab}(x-y)-\alpha \sigma
^2\Delta ^{ab}(x-y)=-\alpha \delta ^{ab}\delta ^4(x-y),  \eqnum{5.8}
\end{equation}
one may find 
\begin{equation}
\begin{array}{c}
\partial _x^\mu \partial _y^\nu \partial _z^\lambda \partial _u^\tau G_{\mu
\nu \lambda \tau }^{abcd}(x,y,z,u)_c=\alpha \mu ^2\{\partial _y^\nu \partial
_z^\lambda G_{~~\nu \lambda }^{dabc}(u,x,y,z)_c \\ 
+\partial _y^\nu \partial _u^\tau G_{~~\nu \tau }^{cabd}(z,x,y,u)_c+\partial
_z^\lambda \partial _u^\tau G_{~~\lambda \tau }^{bacd}(y,x,z,u)_c\}.
\end{array}
\eqnum{5.9}
\end{equation}
This is the W-T identity satisfied by the connected four-point Green
functions. In the Landau gauge, we have 
\begin{equation}
\partial _x^\mu \partial _y^\nu \partial _z^\lambda \partial _u^\tau G_{\mu
\nu \lambda \tau }^{abcd}(x,y,z,u)_c=0  \eqnum{5.10}
\end{equation}
which shows the transversity of the Green function.

The W-T identity for the four-line gluon proper vertex may be derived from
Eq. (5.9) with the help of the following one-particle -irreducible
decompositions of the connected Green functions which may be found by the
standard procedure$^{[11,12]}$, 
\begin{equation}
\begin{array}{c}
G_{\mu \nu \lambda \tau }^{abcd}(x_1,x_2,x_3,x_4)_c \\ 
=\int \prod\limits_{i=1}^4d^4y_iD_{\mu \mu ^{\prime }}^{aa^{\prime
}}(x_1-y_1)D_{\nu \nu ^{\prime }}^{bb^{\prime }}(x_2-y_2)\Gamma _{a^{\prime
}b^{\prime }c^{\prime }d^{\prime }}^{\mu ^{\prime }\nu ^{\prime }\lambda
^{\prime }\tau ^{\prime }}(y_1,y_2,y_3,y_4) \\ 
\times D_{\lambda ^{\prime }\lambda }^{c^{\prime }c}(y_3-x_3)D_{\tau
^{\prime }\tau }^{d^{\prime }d}(y_4-x_4) \\ 
+i\int \prod\limits_{i=1}^4d^4y_id^4z_i\{D_{\mu \mu ^{\prime }}^{aa^{\prime
}}(x_1-y_1)D_{\nu \nu ^{\prime }}^{bb^{\prime }}(x_2-y_2)\Gamma _{a^{\prime
}b^{\prime }e}^{\mu ^{\prime }\nu ^{\prime }\rho }(y_1,y_2,y_3) \\ 
\times D_{\rho \rho ^{\prime }}^{ee^{\prime }}(y_3-z_1)\Gamma _{e^{\prime
}c^{\prime }d^{\prime }}^{\rho ^{\prime }\lambda ^{\prime }\tau ^{\prime
}}(z_1,z_2,z_3)D_{\lambda ^{\prime }\lambda }^{c^{\prime }c}(z_2-x_3)D_{\tau
^{\prime }\tau }^{d^{\prime }d}(z_3-x_4) \\ 
+D_{\mu \mu ^{\prime }}^{aa^{\prime }}(x_1-y_1)D_{\lambda \lambda ^{\prime
}}^{cc^{\prime }}(x_3-y_2)\Gamma _{a^{\prime }c^{\prime }e}^{\mu ^{\prime
}\lambda ^{\prime }\rho }(y_1,y_2,y_3)D_{\rho \rho ^{\prime }}^{ee^{\prime
}}(y_3-z_1) \\ 
\times \Gamma _{e^{\prime }b^{\prime }d^{\prime }}^{\rho ^{\prime }\nu
^{\prime }\tau ^{\prime }}(z_1,z_2,z_3)D_{\nu ^{\prime }\nu }^{b^{\prime
}b}(z_2-x_2)D_{\tau ^{\prime }\tau }^{d^{\prime }d}(z_3-x_4) \\ 
+D_{\nu \nu ^{\prime }}^{bb^{\prime }}(x_2-y_1)D_{\lambda \lambda ^{\prime
}}^{cc^{\prime }}(x_3-y_2)\Gamma _{b^{\prime }c^{\prime }e}^{\nu ^{\prime
}\lambda ^{\prime }\rho }(y_1,y_2,y_3)D_{\rho \rho ^{\prime }}^{ee^{\prime
}}(y_3-z_1) \\ 
\times \Gamma _{e^{\prime }a^{\prime }d^{\prime }}^{\rho ^{\prime }\mu
^{\prime }\tau ^{\prime }}(z_1,z_2,z_3)D_{\mu ^{\prime }\mu }^{a^{\prime
}a}(z_2-x_1)D_{\tau ^{\prime }\tau }^{d^{\prime }d}(z_3-x_4)\}
\end{array}
\eqnum{5.11}
\end{equation}
and 
\begin{equation}
\begin{array}{c}
~G_{~~\lambda \tau }^{abcd}(x_1,x_2,x_3,x_4)_c \\ 
=-i\int \prod\limits_{i=1}^4d^4y_i\Delta ^{aa^{\prime }}(x_1-y_1)\Gamma
_{a^{\prime }b^{\prime }c^{\prime }d^{\prime }}^{~~~~\lambda ^{\prime }\tau
^{\prime }}(y_1,y_2,y_3,y_4)\Delta ^{b^{\prime }b}(y_2-x_2) \\ 
\times D_{\lambda ^{\prime }\lambda }^{c^{\prime }c}(y_3-x_3)D_{\tau
^{\prime }\tau }^{d^{\prime }d}(y_4-x_4) \\ 
+i\int \prod\limits_{i=1}^4d^4y_id^4z_i\{\Delta ^{aa^{\prime
}}(x_1-y_1)\Gamma _{a^{\prime }ed^{\prime }}^{~~~\tau ^{\prime
}}(y_1,y_2,y_3)\Delta ^{ee^{\prime }}(y_2-z_1) \\ 
\times D_{\tau ^{\prime }\tau }^{d^{\prime }d}(y_3-x_4)\Gamma _{e^{\prime
}b^{\prime }c^{\prime }}^{~~~~\lambda ^{\prime }}(z_1,z_2,z_3)\Delta
^{b^{\prime }b}(z_2-x_2)D_{\lambda ^{\prime }\lambda }^{c^{\prime
}c}(z_3-x_3) \\ 
+\Delta ^{aa^{\prime }}(x_1-y_1)\Gamma _{a^{\prime }ec^{\prime
}}^{~~~\lambda ^{\prime }}(y_1,y_2,y_3)\Delta ^{ee^{\prime
}}(y_2-z_1)D_{\lambda ^{\prime }\lambda }^{c^{\prime }c}(y_3-x_3) \\ 
\times \Gamma _{e^{\prime }b^{\prime }d^{\prime }}^{~~~\tau ^{\prime
}}(z_1,z_2,z_3)\Delta ^{b^{\prime }b}(z_2-x_2)D_{\tau ^{\prime }\tau
}^{d^{\prime }d}(z_3-x_4) \\ 
+\Delta ^{aa^{\prime }}(x_1-y_1)\Gamma _{a^{\prime }b^{\prime }e}^{~~~\rho
}(y_1,y_2,y_3)\Delta ^{b^{\prime }b}(y_2-x_2)D_{\rho \rho ^{\prime
}}^{ee^{\prime }}(y_3-z_1) \\ 
\times \Gamma _{e^{\prime }c^{\prime }d^{\prime }}^{\rho ^{\prime }\lambda
^{\prime }\tau ^{\prime }}(z_1,z_2,z_3)D_{\lambda ^{\prime }\lambda
}^{c^{\prime }c}(z_2-x_3)D_{\tau ^{\prime }\tau }^{d^{\prime }d}(z_3-x_4)\}
\end{array}
\eqnum{5.12}
\end{equation}
where $\Gamma _{\mu \nu \lambda \tau }^{abcd}(x_1,x_2,x_3,x_4)$ is the
four-line gluon proper vertex and $\Gamma _{~~\lambda \tau
}^{abcd}(x_1,x_2,x_3,x_4)$ is the four-line ghost-gluon proper vertex. They
are defined as $^{4,5}$%
\begin{equation}
\begin{array}{c}
\Gamma _{\mu \nu \lambda \tau }^{abcd}(x_1,x_2,x_3,x_4)=i\frac{\delta
^4\Gamma }{\delta A^{a\mu }(x_1)\delta A^{b\nu }(x_2)\delta A^{c\lambda
}(x_3)\delta A^{d\tau }(x_4)}|_{J=0}, \\ 
\Gamma _{~~\lambda \tau }^{abcd}(x_1,x_2,x_3,x_4)=\frac{\delta ^4\Gamma }{%
i\delta \bar C^a(x_1)\delta C^b(x_2)\delta A^{c\lambda }(x_3)\delta A^{d\tau
}(x_4)}|_{J=0}.
\end{array}
\eqnum{5.13}
\end{equation}
When substituting Eqs. (5.11) and (5.12) into Eq. (5.9) and transforming Eq.
(5.9) into the momentum space, one can find the following identity satisfied
by the gluon four-line proper vertex 
\begin{equation}
\begin{array}{c}
k_1^\mu k_2^\nu k_3^\lambda k_4^\tau \Lambda _{\mu \nu \lambda \tau
}^{abcd}(k_1,k_2,k_3,k_4)=\Psi \left( 
\begin{array}{cccc}
a & b & c & d \\ 
k_1 & k_2 & k_3 & k_4
\end{array}
\right) \\ 
+\Psi \left( 
\begin{array}{cccc}
a & c & d & b \\ 
k_1 & k_3 & k_4 & k_2
\end{array}
\right) +\Psi \left( 
\begin{array}{cccc}
a & d & b & c \\ 
k_1 & k_4 & k_2 & k_3
\end{array}
\right)
\end{array}
\eqnum{5.14}
\end{equation}
where 
\begin{equation}
\begin{array}{c}
\Psi \left( 
\begin{array}{cccc}
a & b & c & d \\ 
k_1 & k_2 & k_3 & k_4
\end{array}
\right) \\ 
=-ik_1^\mu k_2^\nu \Lambda _{\mu \nu \sigma
}^{abe}(k_1,k_2,-(k_1+k_2))D_{ef}^{\sigma \rho }(k_1+k_2)k_3^\lambda
k_4^\tau \Lambda _{\rho \lambda \tau }^{fcd}(-(k_3+k_4),k_3,k_4) \\ 
+\frac{i\mu ^2}\alpha \chi (k_1^2)\chi (k_2^2)[ik_3^\lambda k_4^\tau \Lambda
_{~~\lambda \tau }^{bacd}(k_2,k_1,k_3,k_4) \\ 
-\Lambda _{~~\sigma }^{bae}(k_2,k_1,-(k_1+k_2))D_{ef}^{\sigma \rho
}(k_1+k_2)k_3^\lambda k_4^\tau \Lambda _{\rho \lambda \tau
}^{fcd}(-(k_3+k_4),k_3,k_4) \\ 
-k_4^\tau \Lambda _{~~\tau }^{bed}(k_2,-(k_2+k_4),k_4)\Delta
^{ef}(k_2+k_4)k_3^\lambda \Lambda _{~~\lambda }^{fac}(-(k_1+k_3),k_1,k_3) \\ 
-k_3^\lambda \Lambda _{~~\lambda }^{bec}(k_2,-(k_2+k_3),k_3)\Delta
^{ef}(k_2+k_3)k_4^\tau \Lambda _{~~~\tau }^{fad}(-(k_1+k_4),k_1,k_4)].
\end{array}
\eqnum{5.15}
\end{equation}
The second and third terms in Eq .(5.14) can be written out from Eq. (5.15)
through cyclic permutations. In the above, we have defined 
\begin{equation}
\begin{array}{c}
\Gamma _{\mu \nu \lambda \tau }^{abcd}(k_1,k_2,k_3,k_4)=(2\pi )^4\delta
^4(\sum_{i=1}^4k_i)\Lambda _{\mu \nu \lambda \tau }^{abcd}(k_1,k_2,k_3,k_4),
\\ 
\Gamma _{~~\lambda \tau }^{abcd}(k_1,k_2,k_3,k_4)=(2\pi )^4\delta
^4(\sum_{i=1}^4k_i)\Lambda _{~~\lambda \tau }^{abcd}(k_1,k_2,k_3,k_4).
\end{array}
\eqnum{5.16}
\end{equation}
In the lowest order approximation, we have checked that except for the first
term in Eq. (5.15) which was encountered in the massless theory, the
remaining mass-dependent terms are cancelled out with the corresponding
terms contained in the second and third terms in Eq. (5.14). Therefore, the
identity in Eq. (5.14) leads to a result in the lowest order approximation
which is consistent with the Feynman rule.

The renormalization of the four-line vertices is similar to that for the
three-line vertices. From the definitions given in Eqs. (5.13), (5.16) and
(4.20), it is clearly seen that the four-line vertices should be
renormalized in such a manner 
\begin{equation}
\begin{array}{c}
\Lambda _{\mu \nu \lambda \tau }^{abcd}(k_1,k_2,k_3,k_4)=Z_3^{-2}\Lambda
_{R\mu \nu \lambda \tau }^{~~abcd}(k_1,k_2,k_3,k_4), \\ 
\Lambda _{~~\lambda \tau }^{abcd}(k_1,k_2,k_3,k_4)=\tilde Z%
_3^{-1}Z_3^{-1}\Lambda _{R\text{ }~~\lambda \tau }^{~~abcd}(k_1,k_2,k_3,k_4).
\end{array}
\eqnum{5.17}
\end{equation}
On inserting these relations into Eqs. (5.14) and (5.15), one can obtain a
renormalized identity similar to Eq. (4.22), that is 
\begin{equation}
\begin{array}{c}
k_1^\mu k_2^\nu k_3^\lambda k_4^\tau \Lambda _{R\mu \nu \lambda \tau }^{%
\text{ }abcd}(k_1,k_2,k_3,k_4)=\Psi _R\left( 
\begin{array}{cccc}
a & b & c & d \\ 
k_1 & k_2 & k_3 & k_4
\end{array}
\right)  \\ 
+\Psi _R\left( 
\begin{array}{cccc}
a & c & d & b \\ 
k_1 & k_3 & k_4 & k_2
\end{array}
\right) +\Psi _R\left( 
\begin{array}{cccc}
a & d & b & c \\ 
k_1 & k_4 & k_2 & k_3
\end{array}
\right) 
\end{array}
\eqnum{5.18}
\end{equation}
where 
\begin{equation}
\begin{array}{c}
\Psi _R\left( 
\begin{array}{cccc}
a & b & c & d \\ 
k_1 & k_2 & k_3 & k_4
\end{array}
\right)  \\ 
=-ik_1^\mu k_2^\nu \Lambda _{R\mu \nu \sigma }^{\text{ }%
abe}(k_1,k_2,-(k_1+k_2))D_{Ref}^{\text{ }\sigma \rho }(k_1+k_2)k_3^\lambda
k_4^\tau \Lambda _{R\rho \lambda \tau }^{fcd}(-(k_3+k_4),k_3,k_4) \\ 
+\frac{i\mu _R^2}{\alpha _R}\chi _R(k_1^2)\chi _R(k_2^2)[ik_3^\lambda
k_4^\tau \Lambda _{R~~\lambda \tau }^{bacd}(k_2,k_1,k_3,k_4) \\ 
-\Lambda _{R~\sigma }^{bae}(k_2,k_1,-(k_1+k_2))D_{Ref}^{\text{ }\sigma \rho
}(k_1+k_2)k_3^\lambda k_4^\tau \Lambda _{R\rho \lambda \tau }^{\text{ }%
fcd}(-(k_3+k_4),k_3,k_4) \\ 
-k_4^\tau \Lambda _{R~~\tau }^{bed}(k_2,-(k_2+k_4),k_4)\Delta
_R^{ef}(k_2+k_4)k_3^\lambda \Lambda _{R~~\lambda }^{\text{ }%
fac}(-(k_1+k_3),k_1,k_3) \\ 
-k_3^\lambda \Lambda _{R~\lambda }^{bec}(k_2,-(k_2+k_3),k_3)\Delta
_R^{ef}(k_2+k_3)k_4^\tau \Lambda _{R~\tau }^{fad}(-(k_1+k_4),k_1,k_4)]
\end{array}
\eqnum{5.19}
\end{equation}
We can also define the vertices $\tilde \Lambda _{\mu \nu \lambda \tau
}^{abcd}(k_1,k_2,k_3,k_4)$ and $\tilde \Lambda _{~~\text{ }\lambda \tau
}^{abcd}(k_1,k_2,k_3,k_4)$ by taking out the coupling constant squared from
the vertices $\Lambda _{\mu \nu \lambda \tau }^{abcd}(k_1,k_2,k_3,k_4)$ and $%
\Lambda _{~~\lambda \tau }^{abcd}(k_1,k_2,k_3,k_4)$, respectively. The
renormalization of these vertices are usually defined by$^{4,5}$%
\begin{equation}
\begin{array}{c}
\tilde \Lambda _{\mu \nu \lambda \tau }^{abcd}(k_1,k_2,k_3,k_4)=Z_4^{-1}%
\tilde \Lambda _{R\mu \nu \lambda \tau }^{~~abcd}(k_1,k_2,k_3,k_4) \\ 
\tilde \Lambda _{~~\lambda \tau }^{abcd}(k_1,k_2,k_3,k_4)=\tilde Z_4^{-1}%
\tilde \Lambda _{R~~\lambda \tau }^{~~abcd}(k_1,k_2,k_3,k_4)
\end{array}
\eqnum{5.20}
\end{equation}
where $Z_4$ and $\tilde Z_4$ are the renormalization constants of the
four-line gluon and ghost-gluon vertices respectively. Obviously, the
identity in Eqs. (5.14) and (5.15) remains formally unchanged if we replace
all the vertices $\Lambda _i$ in the identity with the ones $\widetilde{%
\Lambda }_i$ . Substituting Eqs. (4.27) and (5.20) and the following
renormalization relations for the gluon and ghost particle propagators
(which were given in section 4 of paper I) 
\[
i\Delta ^{ab}(k)=\tilde Z_3i\Delta _R^{ab}(k)
\]
\begin{equation}
iD_{\mu \nu }^{ab}(k)=Z_3iD_{R\mu \nu }^{~~ab}(k)  \eqnum{5.21}
\end{equation}
into such an identity, one may write a renormalized identity similar to Eq.
(4.28), that is 
\begin{equation}
\begin{array}{c}
k_1^\mu k_2^\nu k_3^\lambda k_4^\tau \widetilde{\Lambda }_{R\mu \nu \lambda
\tau }^{\text{ }abcd}(k_1,k_2,k_3,k_4)=\widetilde{\Psi }_R\left( 
\begin{array}{cccc}
a & b & c & d \\ 
k_1 & k_2 & k_3 & k_4
\end{array}
\right)  \\ 
+\widetilde{\Psi }_R\left( 
\begin{array}{cccc}
a & c & d & b \\ 
k_1 & k_3 & k_4 & k_2
\end{array}
\right) +\widetilde{\Psi }_R\left( 
\begin{array}{cccc}
a & d & b & c \\ 
k_1 & k_4 & k_2 & k_3
\end{array}
\right) 
\end{array}
\eqnum{5.22}
\end{equation}
where 
\begin{equation}
\begin{array}{c}
\widetilde{\Psi }_R\left( 
\begin{array}{cccc}
a & b & c & d \\ 
k_1 & k_2 & k_3 & k_4
\end{array}
\right)  \\ 
=\frac{Z_4Z_3}{Z_1^2}\{-ik_1^\mu k_2^\nu \widetilde{\Lambda }_{R\mu \nu
\sigma }^{\text{ }abe}(k_1,k_2,-(k_1+k_2))D_{Ref}^{\text{ }\sigma \rho
}(k_1+k_2)k_3^\lambda k_4^\tau \widetilde{\Lambda }_{R\rho \lambda \tau }^{%
\text{ }fcd}(-(k_3+k_4),k_3,k_4)\} \\ 
+\frac{i\mu _R^2}{\alpha _R}\chi _R(k_1^2)\chi _R(k_2^2)\{i\frac{\widetilde{Z%
}_3Z_4}{Z_3\tilde Z_4}k_3^\lambda k_4^\tau \widetilde{\Lambda }_{R~~\lambda
\tau }^{\text{ }bacd}(k_2,k_1,k_3,k_4) \\ 
-\frac{Z_4\widetilde{Z}_3}{Z_1\tilde Z_1}\widetilde{\Lambda }_{R~~\sigma }^{%
\text{ }bae}(k_2,k_1,-(k_1+k_2))D_{Ref}^{\text{ }\sigma \rho
}(k_1+k_2)k_3^\lambda k_4^\tau \widetilde{\Lambda }_{R\rho \lambda \tau }^{%
\text{ }fcd}(-(k_3+k_4),k_3,k_4) \\ 
-\frac{Z_4\widetilde{Z}_3^2}{Z_3\widetilde{Z}_1^2}[k_4^\tau \widetilde{%
\Lambda }_{R~\tau }^{bed}(k_2,-(k_2+k_4),k_4)\Delta
_R^{ef}(k_2+k_4)k_3^\lambda \widetilde{\Lambda }_{R~~\lambda }^{\text{ }%
fac}(-(k_1+k_3),k_1,k_3) \\ 
+k_3^\lambda \widetilde{\Lambda }_{R~\lambda
}^{bec}(k_2,-(k_2+k_3),k_3)\Delta _R^{ef}(k_2+k_3)k_4^\tau \widetilde{%
\Lambda }_{R~\tau }^{fad}(-(k_1+k_4),k_1,k_4)]\}
\end{array}
\eqnum{5.23}
\end{equation}
Multiplying the both sides of Eqs. (5.22) and (5.23) by $g_R^2$, according
to the relations given in Eqs. (4.29) and in the following 
\begin{equation}
\begin{array}{c}
\Lambda _{R\mu \nu \lambda \tau }^{abcd}(k_{1,}k_2,k_3,k_4)=g_R^2\tilde 
\Lambda _{R\mu \nu \lambda \tau }^{~~abcd}(k_1,k_2,k_3,k_4) \\ 
\Lambda _{R~~\lambda \tau }^{abcd}(k_1,k_2,k_3,k_4)=g_R^2\widetilde{\Lambda }%
_{R~~\lambda \tau }^{~~abcd}(k_1,k_2,k_3,k_4)
\end{array}
\eqnum{5.24}
\end{equation}
we have an identity which is of the same form as the identity in Eqs. (5.22)
and (5.23) except that the vertices $\widetilde{\Lambda }_R^i$ in Eqs.
(5.22) and (5.23) are all replaced by the vertices $\Lambda _R^i$. Comparing
this identity with that written in Eqs. (5.18) and (5.19), one may find 
\begin{equation}
\frac{Z_3Z_4}{Z_1^2}=1,\frac{\widetilde{Z}_3Z_4}{Z_3\widetilde{Z}_4}=1,\frac{%
Z_4\widetilde{Z}_3}{Z_1\widetilde{Z}_1}=1,\frac{Z_4\widetilde{Z}_3^2}{Z_3%
\widetilde{Z}_1^2}=1  \eqnum{5.25}
\end{equation}
which lead to 
\begin{equation}
\frac{Z_1}{Z_3}=\frac{\widetilde{Z}_1}{\widetilde{Z}_3}=\frac{Z_4}{Z_1},%
\frac{Z_1}{\widetilde{Z}_1}=\frac{Z_3}{\widetilde{Z}_3}=\frac{Z_4}{%
\widetilde{Z}_4}.  \eqnum{5.26}
\end{equation}
This just is the S-T identity which is consistent with that given in Eq.
(3.14).

\section{Effective coupling constant}

To concretely demonstrate the renormalizability of the massive gauge field
theory, in this section, we take one-loop renormalization of the QCD with
massive gluons as an example. The renormalization is carried out by the
renormalization group approach$^{[22-24]}$ and gives an exact one-loop
effective coupling constant of the QCD. As argued in our previous paper$%
^{[25-27]}$, when the renormalization is carried out in the mass-dependent
momentum space subtraction scheme$^{[28-31]}$, the solutions to the
renormalization group equations (RGE) satisfied by renormalized wave
functions, propagators and vertices can be uniquely determined by boundary
conditions of the renormalized wave functions, propagators and vertices. In
this case, an exact S-matrix element can be written in the form as given in
the tree-diagram approximation provided that the coupling constant and
particle masses in the matrix element are replaced by their effective
(running) ones which are given by solving their renormalization group
equations. Therefore, the task of renormalization is reduced to find the
solutions of the RGEs for the renormalized coupling constant and particle
masses. Suppose $F_R$ is a renormalized quantity. In the multiplicative
renormalization, it is related to the unrenormalized one $F$ in such a way 
\begin{eqnarray}
F=Z_FF_R  \eqnum{6.1}
\end{eqnarray}
where $Z_F$ is the renormalization constant of $F$. The $Z_F$ and $F_R$ are
all functions of the renormalization point $\mu =\mu _0e^t$ where $\mu _0$
is a fixed renormalization point corresponding the zero value of the group
parameter $t$. Differentiating Eq. (6.1) with respect to $\mu $ and noticing
that the $F$ is independent of $\mu $, we immediately obtain a
renormalization group equation (RGE) satisfied by the function $%
F_R^{[22-24]} $ 
\begin{eqnarray}
\mu \frac{dF_R}{d\mu }+\gamma _FF_R=0  \eqnum{6.2}
\end{eqnarray}
where $\gamma _F$ is the anomalous dimension defined by 
\begin{eqnarray}
\gamma _F=\mu \frac d{d\mu }\ln Z_F.  \eqnum{6.3}
\end{eqnarray}
Since the renormalization constant is dimensionless, the anomalous dimension
can only depend on the ratio ${\beta =\frac{m_R}\mu }${\ where }$m_R$
denotes a, renormalized mass and ${\gamma }_F{=\gamma }_F{(g}_R{,\beta )}$
in which $g_R$ is the renormalized coupling constant and depends on $\mu $.
Since the renormalization point is a momentum taken to subtract the
divergence, we may set $\mu =\mu _0\lambda $ where $\lambda =e^t$ which will
be taken to be the same as in the scaling transformation of momentum $%
p=p_0\lambda $. In the above, $\mu _0$ and $p_0$ are the fixed
renormalization point and momentum respectively. When we set $F$ to be the
coupling constant $g$ and noticing $\mu \frac d{d\mu }=\lambda \frac d{%
d\lambda }$, one can write from Eq. (6.2) the RGE for the renormalized
coupling constant 
\begin{equation}
\lambda \frac{dg_R(\lambda )}{d\lambda }+\gamma _g(\lambda )g_R(\lambda )=0 
\eqnum{6.4}
\end{equation}
with 
\begin{equation}
\gamma _g=\mu \frac d{d\mu }\ln Z_g.  \eqnum{6.5}
\end{equation}
The renormalization constant $Z_g$ is commonly defined by (see Eq. (3.15)) 
\begin{equation}
Z_g=\frac{Z_1}{Z_3^{3/2}}=\frac{\widetilde{Z}_1}{\widetilde{Z_3}Z_3^{\frac 12%
}}  \eqnum{6.6}
\end{equation}
where the last equality is given by using the identity in Eq. (4.31). Here
we would like to choose the expression of $Z_g$ given by the last equality
in Eq. (6.6) to evaluate the anomalous dimension $\gamma _g(\lambda )$. As
denoted in Eqs. (4.25) in paper I and Eq. (4.27), the renormalization
constants $Z_3$, $\widetilde{Z_3}$ and $\widetilde{Z}_1$ are determined by
the gluon self-energy, the ghost article self-energy and the ghost vertex
correction, respectively. At one-loop level, the gluon self-energy is
depicted in Figs. (1a)-(1d), the ghost article self-energy is shown in Fig.
(2) and the ghost vertex correction is represented in Figs. (3a) and (3b).
According to the Feynman rules which are the same as those for the massless
QCD$^{[10]}$ except that the gluon propagator and the ghost particle one are
now given in Eqs. (4.14) and (4.10) in paper I, the expressions of the
self-energies and the vertex correction are easily written out. For the
gluon one-loop self-energy denoted by $-i\Pi _{\mu \nu }^{ab}(k)$, one can
write 
\begin{equation}
\Pi _{\mu \nu }^{ab}(k)=\sum_{i=1}^4\Pi _{\mu \nu }^{(i)ab}(k)  \eqnum{6.7}
\end{equation}
where $\Pi _{\mu \nu )}^{(1)ab}(k)-\Pi _{\mu \nu )}^{(4)ab}(k)$ represent
the self-energies given in turn by Figs.(1a)-(1d). They are separately
represented in the following: 
\begin{equation}
\begin{array}{c}
\Pi _{\mu \nu }^{(1)ab}(k)=i\delta ^{ab}\frac 32g^2\int \frac{d^4l}{(2\pi )^4%
}\frac{g^{\lambda \lambda ^{\prime }}g^{\rho \rho ^{\prime }}}{%
[l^2-m^2+i\varepsilon ][(l+k)^2-m^2+i\varepsilon ]}[g_{\mu \lambda
}(l+2k)_\rho -g_{\lambda \rho }(2l+k)_\mu \\ 
+g_{\rho \mu }(l-k)_\lambda ][g_{\nu \rho ^{\prime }}(l-k)_{\lambda ^{\prime
}}-g_{\lambda ^{\prime }\rho ^{\prime }}(2l+k)_\nu +g_{\lambda ^{^{\prime
}}\nu }(l+2k)_{\rho ^{\prime }}],
\end{array}
\eqnum{6.8}
\end{equation}
\begin{equation}
\Pi _{\mu \nu }^{(2)ab}(k)=-i\delta ^{ab}3g^2\int \frac{d^4l}{(2\pi )^4}%
\frac{(l+k)_\mu l_\nu }{[(l+k)^2-m^2+i\varepsilon ][l^2-m^2+i\varepsilon ]},
\eqnum{6.9}
\end{equation}
\begin{equation}
\Pi _{\mu \nu }^{(3)ab}(k)=-i\delta ^{ab}3g^2\int \frac{d^4l}{(2\pi )^4}%
\frac{g^{\lambda \rho }}{(l^2-m^2+i\varepsilon )}(g_{\mu \nu }g_{\lambda
\rho }-g_{\mu \rho }g_{\lambda \nu })  \eqnum{6.10}
\end{equation}
and 
\begin{equation}
\begin{array}{c}
\Pi _{\mu \nu }^{(4)ab}(k)=-i\delta ^{ab}\frac 12g^2\int \frac{d^4l}{(2\pi
)^4}\frac 1{[(l-k)^2-M^2+i\varepsilon ][l^2-M^2+i\varepsilon ]} \\ 
\times Tr[\gamma _\mu ({\bf l}-{\bf k}+M)\gamma _\nu ({\bf l}+M)]
\end{array}
\eqnum{6.11}
\end{equation}
where ${\bf l=}\gamma ^\lambda l_\lambda $, ${\bf k=}\gamma ^\lambda
k_\lambda $. In the above, $f^{acd}f^{bcd}=3\delta ^{ab}$ and $Tr(T^aT^b)=%
\frac 12\delta ^{ab}$ have been considered. It should be noted that in
writing Eqs. (6.8)-(6.10), we choose to work in the Feynman gauge for
simplicity. This choice is based on the fact that the massive QCD, as proved
in paper I, is an unitary theory, that is to say, the S-matrix elements
evaluated from the massive QCD are independent of gauge parameter.
Therefore, we are allowed to choose a convenient gauge in the calculation.
From Eqs. (6.8)-(6.11), it is clearly seen that 
\begin{equation}
\Pi _{\mu \nu }^{ab}(k)=\delta ^{ab}\Pi _{\mu \nu }(k)=\delta
^{ab}\sum_{i=1}^4\Pi _{\mu \nu }^{(i)}(k).  \eqnum{6.12}
\end{equation}
By the dimensional regularization approach$^{[21]}$, the divergent integrals
over $l$ in Eqs. (6.8)-(6.11) can be regularized in a $n$-dimensional space
and easily calculated. The results are 
\begin{equation}
\begin{array}{c}
\Pi _{\mu \nu }^{(1)}(k)=-\frac 32\frac{g^2}{(4\pi )^2}\int_0^1dx\frac 1{%
\varepsilon [k^2x(x-1)+m^2]^\varepsilon }\{g_{\mu \nu }[11x(x-1) \\ 
+5)k^2+9m^2]+2[5x(x-1)-1]k_\mu k_\nu \},
\end{array}
\eqnum{6.13}
\end{equation}
\begin{equation}
\begin{array}{c}
\Pi _{\mu \nu }^{(2)}(k)=\frac 32\frac{g^2}{(4\pi )^2}\int_0^1dx\frac 1{%
\varepsilon [k^2x(x-1)+m^2]^\varepsilon }\{[k^2x(x-1) \\ 
+m^2]g_{\mu \nu }+2x(x-1)k_\mu k_\nu \},
\end{array}
\eqnum{6.14}
\end{equation}
\begin{equation}
\Pi _{\mu \nu }^{(3)}(k)=\frac{9g^2}{(4\pi )^2}\frac{m^2}\varepsilon g_{\mu
\nu }  \eqnum{6.15}
\end{equation}
and 
\begin{equation}
\Pi _{\mu \nu }^{(4)}(k)=-\frac{4g^2}{(4\pi )^2}\int_0^1dx\frac{k^2x(x-1)}{%
\varepsilon [k^2x(x-1)+M^2]^\varepsilon }[g_{\mu \nu }-\frac{k_\mu k_\nu }{%
k^2}]  \eqnum{6.16}
\end{equation}
where $\varepsilon =2-\frac n2\rightarrow 0$ when $n\rightarrow 4$. In Eqs.
(6.13)-(6.16), except for the $\varepsilon $ in the factor $1/\varepsilon
[k^2x(x-1)+m^2]^\varepsilon $ and $1/\varepsilon [k^2x(x-1)+M^2]^\varepsilon 
$, we have set $\varepsilon \rightarrow 0$ in the other factors and terms by
the consideration that this operation does not affect the calculated result
of the anomalous dimension. According to the decomposition shown in Eqs.
(4.15) and (4.16) in paper I and noticing $g_{\mu \nu }={\cal P}_T^{\mu \nu
}+{\cal P}_L^{\mu \nu }$ where ${\cal P}_T^{\mu \nu }=(g^{\mu \nu }-\frac{%
k^\mu k^\nu }{k^2})$ and ${\cal P}_L^{\mu \nu }=\frac{k^\mu k^\nu }{k^2}$,
it is easy to get the transverse part of $\Pi _{\mu \nu }(k)$ from Eqs.
(6.13)-(6.16) and furthermore, based on the decomposition denoted in Eq.
(4.20) in paper I, i.e., $\Pi _T(k^2)=k^2\Pi _1(k^2)+m^2\Pi _2(k^2)$, the
functions $\Pi _1(k^2)$ and $\Pi _2(k^2)$ can be written out. The results
are 
\begin{equation}
\Pi _1(k^2)=-\frac{g^2}{(4\pi )^2}\int_0^1dx\{\frac{15[2x(x-1)+1]}{%
2\varepsilon [k^2x(x-1)+m^2]^\varepsilon }+\sum\limits_{i=1}^{N_f}\frac{%
4x(x-1)}{\varepsilon [k^2x(x-1)+M_i^2]^\varepsilon }\}  \eqnum{6.17}
\end{equation}
and 
\begin{equation}
\Pi _2(k^2)=-\frac{g^2}{(4\pi )^2}\{\int_0^1dx\frac{12}{\varepsilon
[k^2x(x-1)+m^2]^\varepsilon }-\frac{27}{2\varepsilon m^2}\}.  \eqnum{6.18}
\end{equation}
It is clear that the both functions $\Pi _1(k^2)$ and $\Pi _2(k^2)$ are
divergent in the four-dimensional space-time. When the divergences are
subtracted in the mass-dependent momentum space subtraction scheme$%
^{[28-31]} $, in accordance with the definition in Eq. (4.25) in paper I, we
immediately obtain from the expression in Eq. (6.17) the one-loop
renormalization constant $Z_3$ as follows 
\begin{equation}
\begin{array}{c}
Z_3=1-\Pi _1(\mu ^2) \\ 
=1+\frac{g^2}{(4\pi )^2}\int_0^1dx\{\frac{15[2x(x-1)+1]}{2\varepsilon [\mu
^2x(x-1)+m^2]^\varepsilon }+\sum\limits_{i=1}^{N_f}\frac{4x(x-1)}{%
\varepsilon [\mu ^2x(x-1)+M_i^2]^\varepsilon }\}.
\end{array}
\eqnum{6.19}
\end{equation}

Next, we turn to the ghost particle one-loop self-energy denoted by $%
-i\Omega ^{ab}(q)$. From Fig. (2), in Feynman gauge, one can write 
\begin{equation}
\Omega ^{ab}(q)=i\delta ^{ab}3g^2\int \frac{d^4l}{(2\pi )^4}\frac{q\cdot
(q-l)}{[(q-l)^2-m^2+i\varepsilon ][l^2-m^2+i\varepsilon ]}.  \eqnum{6.20}
\end{equation}
By the dimensional regularization, it is easy to get 
\begin{equation}
\Omega ^{ab}(q)=\delta ^{ab}q^2\hat \Omega (q^2)  \eqnum{6.21}
\end{equation}
where 
\begin{equation}
\hat \Omega (q^2)=\frac{g^2}{(4\pi )^2}\int_0^1dx\frac{3(x-1)}{\varepsilon
[q^2x(x-1)+m^2]^\varepsilon }.  \eqnum{6.22}
\end{equation}
According to the definition given in Eq. (4.25) in paper I and the above
expression , the one-loop renormalization constant of ghost particle
propagator is of the form 
\begin{equation}
\widetilde{Z}_3=1-\hat \Omega (\mu ^2)=1-\frac{g^2}{(4\pi )^2}\int_0^1dx%
\frac{3(x-1)}{\varepsilon [\mu ^2x(x-1)+m^2]^\varepsilon }.  \eqnum{6.23}
\end{equation}

Now, let us discuss the ghost vertex renormalization. In the one-loop
approximation. the vertex defined by extracting out a coupling constant is
expressed as 
\begin{equation}
\widetilde{\Lambda }_\lambda ^{abc}(p,q)=f^{abc}p_\lambda +\Lambda
_{1\lambda }^{abc}(p,q)+\Lambda _{2\lambda }^{abc}(p,q)  \eqnum{6.24}
\end{equation}
where the first term is the bare vertex, the second and the third terms
stand for the one-loop vertex corrections shown in Figs. (3a) and (3b)
respectively. In the Feynman gauge, the vertex corrections are expressed as 
\begin{equation}
\Lambda _{1\lambda }^{abc}(p,q)=-if^{abc}\frac 32g^2\int \frac{d^4l}{(2\pi
)^4}\frac{p\cdot (q-l)(p-l)_\lambda }{[l^2-m^2+i\varepsilon
][(p-l)^2-m^2+i\varepsilon ][(q-l)^2-m^2+i\varepsilon ]}  \eqnum{6.25}
\end{equation}
and 
\begin{equation}
\Lambda _{2\lambda }^{abc}(p,q)=if^{abc}\frac 32g^2\int \frac{d^4l}{(2\pi )^4%
}\frac{l\cdot (p-q-l)p_\lambda -p\cdot lq_\lambda +p\cdot (2q-p+l)l_\lambda 
}{[l^2-m^2+i\varepsilon ][(p-l)^2-m^2+i\varepsilon
][(q-l)^2-m^2+i\varepsilon ]}  \eqnum{6.26}
\end{equation}
where $f^{acd}f^{ebf}f^{dfc}=-\frac 32f^{abc}$ has been noted. By employing
the dimensional regularization to compute the above integrals, it is not
difficult to get 
\begin{equation}
\Lambda _{1\lambda }^{abc}(p,q)=f^{abc}\frac 32\frac{g^2}{(4\pi )^2}%
\int_0^1dx\int_0^1dy\{\frac{\frac 12yp_\lambda }{\varepsilon \Theta
_{xy}^\varepsilon }-\frac 1{\Theta _{xy}}[p_\lambda A_1(p,q)+q_\lambda
B_1(p,q)]-\frac 18p_\lambda \}  \eqnum{6.27}
\end{equation}
where 
\begin{equation}
\begin{array}{c}
\Theta _{xy}=p^2xy(xy-1)+q^2[(x-1)^2y+(x-1)]y-2p\cdot qx(x-1)y^2+m^2, \\ 
A_1(p,q)=\{p\cdot q[1+(x-1)y]-p^2xy\}(1-xy)y, \\ 
B_1(p,q)=\{p\cdot q[1+(x-1)y]-p^2xy\}(x-1)y^2
\end{array}
\eqnum{6.28}
\end{equation}
and 
\begin{equation}
\Lambda _{2\lambda }^{abc}(p,q)=f^{abc}\frac 32\frac{g^2}{(4\pi )^2}%
\int_0^1dx\int_0^1dy\{\frac{\frac 32yp_\lambda }{\varepsilon \Theta
_{xy}^\varepsilon }+\frac 1{\Theta _{xy}}[p_\lambda A_2(p,q)+q_\lambda
B_2(p,q)]-\frac 38p_\lambda \}  \eqnum{6.29}
\end{equation}
where 
\begin{equation}
\begin{array}{c}
A_2(p,q)=\{p^2(2xy-x^2y^2-1)-q^2[(x-1)y-1](x-1)y \\ 
+p\cdot q[2-(3x-2)y+2x(x-1)y^2]\}y, \\ 
B_2(p,q)=[p\cdot q(x-1)-p^2x]y^2.
\end{array}
\eqnum{6.30}
\end{equation}
The divergences in the both vertices $\Lambda _{1\lambda }^{abc}(p,q)$ and $%
\Lambda _{2\lambda }^{abc}(p,q)$ may be subtracted at the renormalization
point $p^2=q^2=\mu ^2$ which implies $k=p-q=0$, being consistent with the
momentum conservation held at the vertices. Upon substituting Eqs. (6.27)
and (6.29) in Eq. (6.24), at the renormalization point, one can get 
\begin{equation}
\widetilde{\Lambda }_\lambda ^{abc}(p,q)\mid _{p^2=q^2=\mu
^2}=f^{abc}p_\lambda (1+\widetilde{L}_1)=\widetilde{Z}_1^{-1}f^{abc}p_\lambda
\eqnum{6.31}
\end{equation}
where 
\begin{equation}
\widetilde{Z}_1=1-\widetilde{L}_1=1-\frac{3g^2}{(4\pi )^2}\int_0^1dx\{\frac x%
{\varepsilon [\mu ^2x(x-1)+m^2]^\varepsilon }-\frac{x^2(x-1)\mu ^2}{\mu
^2x(x-1)+m^2}-\frac 14\}  \eqnum{6.32}
\end{equation}
which is the one-loop renormalization constant of the ghost vertex.

Now we are ready to calculate the anomalous dimension $\gamma _g(\lambda )$.
Substituting the expressions in Eqs. (6.6), (6.19), (6.23) and (6.32) into
Eq. (6.5), it is easy to find an analytical expression of the anomalous
dimension $\gamma _g(\lambda )$. When we set $\frac m\mu =\frac \beta \lambda
$ and $\frac{M_i}\mu =\frac{\rho _i}\lambda $ with defining $\beta =\frac m%
\Lambda $ and $\rho _i=\frac{M_i}\Lambda $ (here we have set $\mu _0\equiv
\Lambda $), the expression of $\gamma _g(\lambda )$, in the approximation of
order $g^2$, is given by 
\begin{equation}
\gamma _g(\lambda )=\lim\limits_{\varepsilon \rightarrow 0}[\mu \frac d{d\mu 
}\ln \widetilde{Z}_1-\mu \frac d{d\mu }\ln \widetilde{Z}_3-\frac 12\mu \frac 
d{d\mu }\ln Z_3]=\frac{g_R^2}{(4\pi )^2}F(\lambda )  \eqnum{6.33}
\end{equation}
where 
\begin{equation}
\begin{array}{c}
F(\lambda )=\frac{19}2-\frac{15\beta ^2}{\lambda ^2}+\frac{3\lambda ^2}{%
2(\lambda ^2-4\beta ^2)}-(8-\frac{10\beta ^2}{\lambda ^2}-\frac{\lambda ^2}{%
\lambda ^2-4\beta ^2})\frac{3\beta ^2}{\lambda \sqrt{\lambda ^2-4\beta ^2}}
\\ 
\times \ln \frac{\lambda -\sqrt{\lambda ^2-4\beta ^2}}{\lambda +\sqrt{%
\lambda ^2-4\beta ^2}}-\frac 23\sum\limits_{i=1}^{N_f}[1+\frac{6\rho _i^2}{%
\lambda ^2}-\frac{12\rho _i^4}{\lambda ^3\sqrt{\lambda ^2-4\rho _i^2}}\ln 
\frac{\lambda -\sqrt{\lambda ^2-4\rho _i^2}}{\lambda +\sqrt{\lambda ^2-4\rho
_i^2}}]
\end{array}
\eqnum{6.34}
\end{equation}
in which $N_f$ denotes the number of quark flavors. We would like to note
that the fixed renormalization point $\Lambda $ in $\beta $ and $\rho _i$
can be taken arbitrarily. For example, the $\Lambda $ may be chosen to be
the mass of the quark of $N_f$ -th flavor. In this case, $\beta =m/M_{N_f}$
and $\rho _i=M_i/M_{N_f}$. In practice, the $\Lambda $ will be treated as a
scaling parameter of renormalization.

With the $\gamma _g(\lambda )$ given above, the equation in Eq. (6.4) can be
solved to give the effective coupling constant as follows 
\begin{equation}
\alpha _R(\lambda )=\frac{\alpha _R}{1+\frac{\alpha _R}{2\pi }G(\lambda )} 
\eqnum{6.35}
\end{equation}
where $\alpha _R=\alpha _R(1)$ and 
\begin{equation}
G(\lambda )=\int_1^\lambda \frac{d\lambda }\lambda =\varphi _1(\lambda
)-\varphi _1(1)-\frac 13\sum\limits_{i=1}^{N_f}[\varphi _2^i(\lambda
)-\varphi _2^i(1)]  \eqnum{6.36}
\end{equation}
in which 
\begin{equation}
\varphi _1(\lambda )=[(19-\frac{10\beta ^2}{\lambda ^2})\frac{\sqrt{\lambda
^2-4\beta ^2}}{4\lambda }+\frac{3\lambda }{4\sqrt{\lambda ^2-4\beta ^2}}]\ln 
\frac{\lambda +\sqrt{\lambda ^2-4\beta ^2}}{\lambda -\sqrt{\lambda ^2-4\beta
^2}}+\frac{5\beta ^2}{\lambda ^2},  \eqnum{6.37}
\end{equation}
\begin{equation}
\varphi _2^i(\lambda )=(1+\frac{2\rho _i^2}{\lambda ^2})\frac{\sqrt{\lambda
^2-4\rho _i^2}}\lambda \ln \frac{\lambda +\sqrt{\lambda ^2-4\rho _i^2}}{%
\lambda -\sqrt{\lambda ^2-4\rho _i^2}}-\frac{4\rho _i^2}{\lambda ^2} 
\eqnum{6.38}
\end{equation}
and $\varphi _1(1)=\varphi _1(\lambda )\mid _{\lambda =1}$, $\varphi
_2^i(1)=\varphi _2^i(\lambda )\mid _{_{\lambda =1}}$. In the large momentum
limit ($\lambda \rightarrow \infty $), we have 
\begin{equation}
G(\lambda )=(11-\frac 23N_f)\ln \lambda .  \eqnum{6.39}
\end{equation}
This just is the result for massless QCD which was obtained previously in
the minimal subtraction scheme$^{[18-20]}$. It should be noted that the
expressions in Eqs. (6.34), (6.37) and (6.38) are obtained at the timelike
subtraction point where the $\lambda $ is a real variable. We may also take
spacelike momentum subtraction. For this kind of subtraction, corresponding
to $\mu \rightarrow i\mu $, the variable $\lambda $ in Eqs. (6.34), (6.37)
and (6.38) should be replaced by $i\lambda $ where $\lambda $ is still a
real variable. It is easy to see that the function in Eq. (6.39) is the same
for the both subtractions.

The behavior of the function $\alpha _R(\lambda )$ is graphically described
in Figs. (4) and (5). Figs. (4) and (5) represent respectively the effective
coupling constants obtained at the timelike subtraction point and the
spacelike subtraction point, where we take the flavor $N_f=3$ as an
illustration. The parameters $\alpha _R,$ $m$ and $\Lambda $ are taken to be 
$\alpha _R=0.2$, $m=600MeV$, $\Lambda =500MeV$. The masses of up, down and
strange quarks are taken to be constituent quark ones, i.e., $M_u=M_d=350MeV$
and $M_s=500MeV$. For comparison, we show in each of the figures three
effective coupling constants which are obtained by the the massive QCD, the
massless QCD and the minimal subtraction, respectively. These effective
coupling constants are respectively represented by the solid, dashed and
dotted lines in the figures. From Fig. (4) it is seen that the effective
coupling constant given by the massive QCD is an analytical function with a
maximum at $\lambda =1.346$, the effective coupling constant given by the
massless QCD is also an analytical function with a peak around $\lambda =1$,
but the effective coupling constant given by the minimal subtraction has a
singularity at $\lambda =$ $0.1746$. Fig. (5) indicates that the effective
coupling constant given by the massive QCD, similar to the one given by the
minimal subtraction, has a Landau pole at $\lambda =0.1845$ which implies
that the coupling constant is not applicable in the region $\lambda \leq
0.1845$. However, if the gluon mass is taken to be $m\leq 425.75MeV$, we
find, the Landau pole disappears and the effective coupling constant,
analogous to the effective coupling constant given by the massless QCD ,
becomes a smooth function in the whole region of momentum as illustrated by
the dotted-dashed line in Fig. (5) which represents the effective coupling
constant given by taking $m=425.75MeV$.

In this paper, we limit ourself to show the derivation and result of the
effective coupling constant only. The effective gluon mass and quark mass
can be derived in the similiar way. All these effective quantities, as
expected, are of asymptotic behaviors.

\section{ Concluding remarks}

The derivation and the result stated in section 2 clearly show that the
divergences appearing in the perturbative calculations for the massive
non-Abelian gauge field theory can indeed be eliminated by introducing a
finite number of counterterms as shown in Eqs. (2.27)-(2.31) and (2.34).
Saying equivalently, as shown in section 3, these divergences may be
absorbed into a finite number of renormalization constants and wave
functions and thus be removed by redefining the wave functions and the
physical parameters. In view of this, according to the general argument of
renormalizability, we can say that the renormalizability of the QCD with
massive gluons is absolutely no problem. The renormalizability of the
massive QCD was illustrated in section 6 where we have derived a rigorous
one-loop effective coupling constant and shown that the corresponding result
obtained in the previous literature$^{[18-20]}$ is only an approximate one
given in the large momentum limit. Thus, contrary to the prevailing concept
that it is impossible to build a renormalizable massive gauge field theory
without recourse to the Higgs mechanism $^{[2-10]}$, we have succeeded in
establishing such a theory which is renormalizable. The basic idea to
achieve this success is the consideration that the massive gauge field must
be viewed as a constrained system in the whole space of the full vector
potential. Therefore, to construct a correct quantum theory for such a
system, the unphysical degrees of freedom contained in the massive
Yang-Mills Lagrangian must be eliminated by introducing appropriate
constraint conditions on the gauge field and the gauge group. In doing this,
the requirement of gauge-invariance must be respected in the whole process
of the construction of the theory. Particularly, in the physical space
defined by the Lorentz condition, only infinitesimal gauge transformations
are necessary to be considered. These essential points were not realized
clearly and handled correctly in the previous studies. In the earlier works
of investigating the massive non-Abelian gauge field theory$^{[2-10]}$, as
mentioned in Introduction, authors all started with the massive Yang-Mills
Lagrangian and considered that this Lagrangian itself forms a complete
description of the massive gauge field dynamics. When using this Lagrangian
to construct the quantum theory, they found that except for the neutral
vector meson field in interaction with a conserved current, the theory is
nonrenormalizable because of the presence of the mass term. However, as
pointed out in Ref.[1], when the Lorentz condition is introduced and thereby
the infinitesimal gauge transformations are taken into account only, the
problem of nonrenormalizability will all disappear in those works.

Another point we would like to emphasize is that for the massive gauge field
theory, the ghost particle acquires a spurious mass $\mu $ in general
gauges. This mass term is necessary to be introduced so as to compensate the
gauge-non-invariance of the gauge boson mass term and preserve the effective
action to be gauge-invariant. As we see, the occurrence of this mass term in
the theory is essential to guarantee the theory to be self-consistent.
Otherwise, for example, if lack of this mass term in the ghost particle
propagator, the W-T identities shown in Eqs. (4.13), (5.14) and (5.15),
which are derived from Eqs. (4.6) and (5.9) respectively, will have
different expressions. In the lowest order approximation, these expressions
can not be converted to the results which coincide with the Feynman rules.
At this point, we can say, some previous works$^{[10]}$ are not correct
because these theories did not give the Lagrangian a ghost particle mass
term in the general gauges.

At present, the massless QCD has widely been recognized to be the candidate
of the strong interaction theory and has been proved to be compatible with
the present experiments. However, we think, the QCD with massive gluons
would be more favorable to explain the strong interaction phenomenon,
particularly, at the low energy domain because the massive gluon would make
the force range more shorter than that caused by the massless gluon. As for
the high energy and large momentum transfer phenomena, as seen from the
massive gluon propagator, the gluon mass gives little influence on the
theoretical result in this case so that the massive QCD could not conflict
with the well-established results gained from the massless QCD in the high
energy region.

\subsection{\bf ACKNOWLEDGMENTS }

We would like to thank Dr. Hai-Jun Wang for numerical calculations and
plotting the figures. This project was supported by National Natural Science
Foundation of China.

\section{Appendix: Verification of the solution to the W-T identity
satisfied by the counterterms}

In this appendix, it is shown that the solution given in Eq. (2.27) indeed
satisfies the W-T identity written in Eq. (2.25) (or in Eq. (2.21)). Since
the $H_\alpha $ in the first term of Eq. (2.27) are the gauge-invariant
functional of the field functions $A_\mu ^a,\bar \psi $ and $\psi $,
according to the definition in Eq. (2.19), it is seen that$^{[10]}$%
\begin{equation}
\begin{array}{c}
\rho (\hat S_0)H_\alpha =\frac{\delta \hat S_0}{\delta \varphi _i}\cdot 
\frac{\delta H_\alpha }{\delta u_i}+\frac{\delta \hat S_0}{\delta u_i}\cdot 
\frac{\delta H_\alpha }{\delta \varphi _i}=\frac{\delta \hat S_0}{\delta u_i}%
\cdot \frac{\delta H_\alpha }{\delta \varphi _i} \\ 
=\Delta \varphi _i\cdot \frac{\delta H_\alpha }{\delta \varphi _i}=\xi
^{-1}\delta \varphi _i\cdot \frac{\delta H_\alpha }{\delta \varphi _i}=\xi
^{-1}\delta H_\alpha =0
\end{array}
\eqnum{A1}
\end{equation}
where the fact that the $H_\alpha $ are independent of the source variables
and gauge-invariant, $\delta H_\alpha =0$, has been noted. For the second
term in Eq. (2.27), it is only needed to prove that the operator $\rho (\hat 
S_0)$ is nilpotent$^{[10]}$. Let us calculate 
\begin{equation}
\begin{array}{c}
\rho (\hat S_0)^2F^n=\frac{\delta \hat S_0}{\delta \varphi _j}\cdot \frac 
\delta {\delta u_j}(\frac{\delta \hat S_0}{\delta \varphi _i}\cdot \frac{%
\delta F_n}{\delta u_i}+\frac{\delta \hat S_0}{\delta u_i}\cdot \frac{\delta
F_n}{\delta \varphi _i}) \\ 
+\frac{\delta \hat S_0}{\delta u_j}\cdot \frac \delta {\delta \varphi _j}(%
\frac{\delta \hat S_0}{\delta \varphi _i}\cdot \frac{\delta F_n}{\delta u_i}+%
\frac{\delta \hat S_0}{\delta u_i}\cdot \frac{\delta F_n}{\delta \varphi _i})
\\ 
=\frac{\delta \hat S_0}{\delta \varphi _j}\cdot (\frac{\delta ^2\hat S_0}{%
\delta u_j\delta \varphi _i}\cdot \frac{\delta F_n}{\delta u_i}+\frac{\delta 
\hat S_0}{\delta \varphi _i}\cdot \frac{\delta ^2F_n}{\delta u_j\delta u_i}
\\ 
+\frac{\delta ^2\hat S_0}{\delta u_j\delta u_i}\cdot \frac{\delta F_n}{%
\delta \varphi _i}-\frac{\delta \hat S_0}{\delta u_i}\cdot \frac{\delta ^2F_n%
}{\delta u_j\delta \varphi _i}) \\ 
+\frac{\delta \hat S_0}{\delta u_j}\cdot (\frac{\delta ^2\hat S_0}{\delta
\varphi _j\delta \varphi _i}\cdot \frac{\delta F_n}{\delta u_i}+\frac{\delta 
\hat S_0}{\delta \varphi _i}\cdot \frac{\delta ^2F_n}{\delta \varphi
_j\delta u_i} \\ 
+\frac{\delta ^2\hat S_0}{\delta \varphi _j\delta u_i}\cdot \frac{\delta F_n%
}{\delta \varphi _i}+\frac{\delta \hat S_0}{\delta u_i}\cdot \frac{\delta
^2F_n}{\delta \varphi _j\delta \varphi _i}) \\ 
=\frac \delta {\delta \varphi _i}(\frac{\delta \hat S_0}{\delta \varphi _j}%
\cdot \frac{\delta \hat S_0}{\delta u_j})\cdot \frac{\delta F_n}{\delta u_i}+%
\frac \delta {\delta u_i}(\frac{\delta \hat S_0}{\delta \varphi _j}\cdot 
\frac{\delta \hat S_0}{\delta u_j})\cdot \frac{\delta F_n}{\delta \varphi _i}
\\ 
+\frac{\delta \hat S_0}{\delta u_j}\cdot \frac{\delta \hat S_0}{\delta
\varphi _i}\cdot \frac{\delta ^2F_n}{\delta \varphi _j\delta u_i}-\frac{%
\delta \hat S_0}{\delta \varphi _j}\cdot \frac{\delta \hat S_0}{\delta u_i}%
\cdot \frac{\delta ^2F_n}{\delta u_j\delta \varphi _i} \\ 
+\frac{\delta \hat S_0}{\delta \varphi _j}\cdot \frac{\delta \hat S_0}{%
\delta \varphi _i}\cdot \frac{\delta ^2F_n}{\delta u_j\delta u_i}+\frac{%
\delta \hat S_0}{\delta u_j}\cdot \frac{\delta \hat S_0}{\delta u_i}\cdot 
\frac{\delta ^2F_n}{\delta \varphi _j\delta \varphi _i}.
\end{array}
\eqnum{A2}
\end{equation}
In the above, we assume that the field variables are commuting and the
source variables are anticommuting for convenience of statement. With this
assumption, It is easy to see that the third and fourth terms are cancelled
with each other and the fifth and sixth terms themselves are identical to
zero because $\frac{\delta \hat S_0}{\delta \varphi _j}\cdot \frac{\delta 
\hat S_0}{\delta \varphi _i}$ and $\frac{\delta ^2F_n}{\delta \varphi
_j\delta \varphi _i}$ are symmetric tensors, while $\frac{\delta ^2F_n}{%
\delta u_j\delta u_i}$ and $\frac{\delta \hat S_0}{\delta u_j}\cdot \frac{%
\delta \hat S_0}{\delta u_i}$ are antisymmetric ones. Therefore, we have 
\begin{equation}
\begin{array}{c}
\rho (\hat S_0)^2F^n=\frac \delta {\delta \varphi _i}(\hat S_0*\hat S%
_0)\cdot \frac{\delta F_n}{\delta u_i}+\frac \delta {\delta u_i}(\hat S_0*%
\hat S_0)\cdot \frac{\delta F_n}{\delta \varphi _i} \\ 
=(\hat S_0*\hat S_0)*F_n=0
\end{array}
\eqnum{A3}
\end{equation}
where $(\hat S_0*\hat S)=0$ has been noted. Similarly, it can be proved 
\begin{equation}
\rho (\hat S_{n-1})^2F_n=0.  \eqnum{A4}
\end{equation}

\section{Reference}

\begin{itemize}
\item[1]  J. C. Su, IL Nuovo Cimento, {\bf 117 B} (2002) 203.

\item[2]  J. Sakurai, Ann. Phys. {\bf 11}, (1960) 1.

\item[3]  M. Gell-Mann, Phys. Rev. {\bf 125} (1962) 1067.

\item[4]  S. L. Glashow and M. Gell-Mann, Ann. Phys. {\bf 15} (1961) 437.

\item[5]  H. Umezawa and S. Kamefuchi, Nucl. Phys. {\bf 23} (1961) 399.

\item[6]  P. A. Ionides, Nucl. Phys. {\bf 23} (1961) 662.

\item[7]  A. Salam, Nucl. Phys. {\bf 18}, (1960) 681; Phys. Rev. {\bf 127}
(1962) 331.

\item[8]  D. G. Boulware, Ann. Phys. {\bf 56} (1970) 140.

\item[9]  A. Salam and J. Strathdee, Phys. Rev. {\bf D 2} (1970) 2869.

\item[10]  C. Itzykson and F-B. Zuber, Quantum Field Theory, McGraw-Hill,
New York (1980).

\item[11]  E. S. Abers and B. W. Lee, Phys. Rep. {\bf C9} (1973) 1; B.
W.Lee, In Methods in Field Theory (1976), ed. R. Balian and Zinn-Justin.

\item[12]  L. D. Faddeev and A. A. Slavnov, Gauge Fields: Introduction to
Quantum Theory, The Benjamin Cummings Publishing Company Inc. (1980).

\item[13]  N. N. Bogoliubov and O. Parasiuk, Acta. Math. {\bf 97} (1957) 227.

\item[14]  K. Hepp, Commun. Math. Phys. {\bf 2} (1966) 301.

\item[15]  W. Zimmermann, in Lectures on Elementary Particles and Quantum
Field Theory, edited by S. Deser et al. (MIT Press, Combride, Mass, (1970).

\item[16]  A. Slavnov, Theor. and Math. Phys. {\bf 10} (1972) 99, (English
translation).

\item[17]  J. C. Taylor, Nucl. Phys. {\bf B 33} (1971) 436.

\item[18]  H. D. Politzer. Phys. Rev. Lett. {\bf 30}, 1346 (1973).

\item[19]  D. J. Gross and F. Wilczek, Phys. Rev. Lett. {\bf 30}, 1343
(1973); Phys. Rev.{\bf \ D} {\bf 8}, 3633 (1973)

\item[20]  H. Georgi and H. D. Politzer, Phys. Rev. {\bf D 14}, 1829 (1976).

\item[21]  G. t' Hooft and M. Veltman, Nucl. Phys. {\bf B 44} (1972) 189.

\item[22]  C. G. Callan, Phys. Rev. {\bf D 2}, 1541 (1970); K. Symanzik,
Commun, Math. Phys. {\bf 18}, 227 (1970).

\item[23]  S. Weinberg, Phys. Rev. {\bf D }8, 3497 (1973).

\item[24]  J. C. Collins and A. J. Macfarlane, Phys. Rev. {\bf D 10}, 1201
(1974).

\item[25]  J. C. Su, X. X. Yi and Y.H. Cao, J. Phys. G: Nucl. Part. Phys. 
{\bf 25}, 2325 (1999).

\item[26]  J. C. Su, L. Shan and Y. H. Cao, Commun. Theor. Phys. {\bf 36},
665 (2000).
\end{itemize}

\begin{itemize}
\item[27]  J. C. Su and Hai-Jun Wang, Phys. Rev. {\bf C 70}, 044003 (2004).

\item[28]  W. A. Bardeen, A. J. Buras, D. W. Duke and T. Muta, Phys. Rev. 
{\bf D} {\bf 18}, 3998 (1978); W. A. Bardeen and R. A. J. Buras, Phys. Rev. 
{\bf D} {\bf 20}, 166 (1979).

\item[29]  W. Celmaster and R. J. Gonsalves, Phys. Rev. Lett. {\bf 42}, 1435
(1979); Phys. Rev. {\bf D} {\bf 20}, 1420 (1979);
\end{itemize}

W. Celmaster and D. Sivers, Phys. Rev. {\bf D} {\bf 23}, 227 (1981).

\begin{itemize}
\item[30]  E. Braaten and J. P. Leveille, Phys. Rev. {\bf D} {\bf 24}, 1369
(1981).

\item[31]  S. N. Gupta and S. F. Radford, Phys. Rev. {\bf D} {\bf 25}, 2690
(1982); J. C. Collins and A. J. Macfarlane, Phys. Rev. {\bf D} {\bf 10},
1201 (1974).
\end{itemize}

\section{Figure captions}

Fig. (1): The one-loop gluon self-energy. The solid, wavy and dashed lines
represent the free quark, gluon and ghost particle propagators respectively.

Fig. (2): The one-loop ghost particle self-energy. The lines represent the
same as in Fig. (1).

Fig. (3): The one-loop ghost-gluon vertices . The lines mark the same as in
Fig. (1).

Fig. (4): The one-loop effective coupling constants ${\alpha _R(\lambda )}$
given by the timelike momentum space subtraction. The solid and dashed lines
represent the coupling constants given by the massive QCD and the massless
QCD, respectively. The dotted line denotes the coupling constant given in {%
the minimal subtraction scheme}.

Fig. (5): The one-loop effective coupling constants ${\alpha _R(\lambda )}$
given by the spacelike momentum space subtraction. The solid and dashed
lines represent the coupling constants given by the massive QCD and the
massless QCD, respectively. The dotted line denotes the coupling constant
given in {the minimal subtraction scheme}. The dashed-dotted line represents
the one-loop effective coupling constants ${\alpha _R(\lambda )}$ given by
the spacelike momentum subtraction for which the gluon mass is taken to be a
smaller value.

\end{document}